\documentclass[]{aa}
\usepackage{graphicx}
\usepackage{natbib}
\usepackage{amsmath}
\usepackage[modulo, switch]{lineno}
\usepackage{txfonts}
\def\mh{\,$\mu$Hz}

\def\teff{$T_{\mathrm{eff}}$}
\def\lg{\ensuremath{\log g}}
\def\p1{Paper\,I}
\def\num{$\nu_\mathrm{max}$}
\def\dnu{$\Delta\nu$}
\def\dn1{$\delta\nu_{01}$}
\def\dn2{$\delta\nu_{02}$}
\def\sun{\hbox{$_\odot$}}

\begin{document}
%
   \title{Asteroseismology of red giants from the first four months of \textit{Kepler} data: Fundamental stellar parameters}

   \author{T. Kallinger\inst{1, 2}
 	  \and
	  B. Mosser\inst{3}
	  \and
	  S. Hekker\inst{4}
	  \and
	  D. Huber\inst{5}
  	  \and
	  D. Stello\inst{5}
	  \and
	  S. Mathur\inst{6}
	  \and
	  S. Basu\inst{7}
	  \and
	  T. R. Bedding\inst{5}
	  \and
	  W. J. Chaplin\inst{4}
	  \and
	  J. De Ridder\inst{8}
	  \and
	  Y. P. Elsworth\inst{4}
	  \and
	  S. Frandsen\inst{9}
	  \and
	  R.A. Garc\'\i a\inst{10}
	  \and
          M. Gruberbauer\inst{11}
          \and
          J. M. Matthews\inst{2}
          \and
          W.J. Borucki\inst{12}
          \and
          H. Bruntt\inst{13}
          \and
          J. Christensen-Dalsgaard\inst{13}
          \and
          R.L. Gilliland\inst{14}
          \and
          H. Kjeldsen\inst{13}
          \and
          D. G. Koch\inst{12}
            }

   \offprints{kallinger@phas.ubc.ca}

   \institute{
Institute for Astronomy (IfA), University of Vienna, T\"urkenschanzstrasse 17, 1180 Vienna, Austria
              \and
Department of Physics and Astronomy, University of British Columbia, 6224 Agricultural Road, Vancouver, BC V6T 1Z1, Canada
		\and 
LESIA, CNRS, Universit\'e Pierre et Marie Curie, Universit\'e Denis, Diderot, Observatoire de Paris, 92195 Meudon cedex, France
		\and		
School of Physics and Astronomy, University of Birmingham, Edgbaston, Birmingham B15 2TT, UK
		\and
Sydney Institute for Astronomy (SIfA), School of Physics, University of Sydney, NSW 2006, Australia
		\and		
High Altitude Observatory, NCAR, P.O. Box 3000, Boulder, CO 80307, USA
		\and		
Department of Astronomy, Yale University, P.O. Box 208101, New Haven, CT 06520-8101, USA
		\and
Instituut voor Sterrenkunde, K.U. Leuven, Celestijnenlaan 200D, 3001 Leuven, Belgium
		\and
Danish AsteroSeismology Centre (DASC), Department of Physics and Astronomy, Aarhus University, DK-8000 Aarhus C, Denmark		
		\and
Laboratoire AIM, CEA/DSM-CNRS, Universit\'e Paris 7 Diderot, IRFU/SAp, Centre de Saclay, 91191, GIf-sur-Yvette, France
		\and
Institute for Computational Astrophysics, Department of Astronomy and Physics, Saint Marys University, Halifax, NS B3H 3C3, Canada
	\and
NASA Ames Research Center, MS 244-30, Moffett Field, CA 94035, USA
	\and
Department of Physics and Astronomy, Building 1520, Aarhus University, 8000 Aarhus C, Denmark             
	\and
Space Telescope Science Institute, 3700 San Martin Drive, Baltimore, MD 21218, USA
}

   \date{Received ; accepted }

\abstract
{Clear power excess in a frequency range typical for solar-type oscillations in red giants has been detected in more than 1\,000 stars, which have been observed during the first 138 days of the science operation of the NASA \textit{Kepler} satellite. This sample includes stars in a wide mass and radius range with spectral types G and K, extending in luminosity from the bottom of the giant branch up to high-luminous red giants, including the red bump and clump. The high-precision asteroseismic observations with \textit{Kepler} provide a perfect source for testing stellar structure and evolutionary models, as well as investigating the stellar population in our Galaxy.}
{We aim to extract accurate seismic parameters from the \textit{Kepler} time series and use them to infer asteroseismic fundamental parameters from scaling relations and a comparison with red-giant models.}
{We fit a global model to the observed power density spectra, which allows us to accurately estimate the granulation background signal and the global oscillation parameters, such as the frequency of maximum oscillation power. We find regular patterns of radial and non-radial oscillation modes and use a new technique to automatically identify the mode degree and the characteristic frequency separations between consecutive modes of the same spherical degree. In most cases, we can also measure the small separation between $l$ = 0, 1, and 2 modes. Subsequently, the seismic parameters are used to estimate stellar masses and radii and to place the stars in an H-R diagram by using an extensive grid of stellar models that covers a wide parameter range. Using Bayesian techniques throughout our entire analysis allows us to determine reliable uncertainties for all parameters.}
{We provide accurate seismic parameters and their uncertainties for a large sample of red giants and determine their asteroseismic fundamental parameters. We investigate the influence of the stars' metallicities on their positions in the H-R diagram. Finally, we study the red-giant populations in the red clump and bump and compare them to a synthetic population. We find a mass and metallicity gradient in the red clump and clear evidence of a secondary-clump population.}
{}

\authorrunning{Kallinger et al.}
\titlerunning{Asteroseismic fundamental parameters of \textit{Kepler} red giants}
   \maketitle

\section{Introduction}	\label{sec:intro}

Studying solar-type oscillations has proved to be a powerful way to test the physical processes in stars \citep[e.g.][]{jcd04} that are similar to our Sun and also to the more evolved red giants, which represent the future of our Sun. The turbulent motions in the convective envelopes of these stars produce an acoustic noise that can stochastically drive (and damp) resonant p-mode oscillations, typically with small amplitudes. On the other hand, the global properties of solar-type oscillations, such as the frequency range where they are excited to observable amplitudes and their characteristic spacings, are predominantly defined by the stellar mass and radius. By using accurate asteroseismic data, it should therefore be possible to constrain fundamental parameters to levels of precision that would otherwise be impossible. This has important applications in, for example, exoplanet studies, which depend on firm knowledge of the fundamental parameters of the host star. Asteroseismic data can put tight constraints on the absolute radii of transiting planets, or determine the age of an exoplanetary system \citep[e.g.][]{jcd10}.

An obvious requirement for such asteroseismic studies is the availability of accurate observational data. The first indications of solar-type oscillations in G and K-type giants were based on ground-based observations in radial velocity (Arcturus: \citealt{mer99}; $\xi$\,Hya: \citealt{fra02}; $\epsilon$\,Oph: \citealt{rid06}) and photometry (M67: \citealt{ste07}), which largely suffered from low signal-to-noise data sets and aliasing. The periods of solar-type oscillations in red giants range from hours to days and hence call for long and preferably uninterrupted observations to resolve the oscillations, which can be done best from space. Space-based detection were made with the star tracker of the \textit{Wide Field Infrared Explorer} satellite \citep[WIRE; e.g.][]{buz00,ret03}, the \textit{Hubble Space Telescope} \citep[HST; e.g.][]{edm96, kal05, ste09a}, \textit{Microvariability and Oscillation of Stars} \citep[MOST;][]{bar07, kal08a, kal08b}, and the \textit{Solar Mass Ejection Imager} \citep[SMEI;][]{tar07}. Significant improvement in quality and quantity of the observations came from the 150-day long observations with the \textit{Convection, Rotation and planetary Transits} satellite (CoRoT), which provided clear detections of radial and non-radial oscillation modes in numerous stars \citep{rid09,hek09,car10,kal10, mos10}. Most recently, the NASA \textit{Kepler Mission} has demonstrated its great asteroseismic potential to observe solar-type oscillations in red giants \citep{bed10,hek10,ste10}. We refer to our companion papers presenting a more detailed study of the asteroseismic observables \citep{hub10} and a comparison of global oscillation parameters derived using different methods \citep{hek10b}. 

The oscillation spectrum of a solar-type oscillating star presents a pattern of modes with nearly regular frequency spacings, where the signature of these spacings carries information about the internal structure of the star. The large frequency spacing (\dnu ), for example, is the frequency differences between consecutive overtones having the same spherical degree ($l$), and is related to the acoustic radius and therefore to the mean density of the star \citep{bro91,kje95}. Another directly accessible seismic parameter, the frequency of maximum oscillation power (\num ), is related to the acoustic cut-off frequency and therefore, in the adiabatic case and under the assumption of an ideal gas, defined by the surface gravity and effective temperature of the star \citep[e.g.][]{kje95}. 

Estimating fundamental parameters from these seismic parameters has become an important application of asteroseismic observations. Recent investigations in this context were made by \citet{ste08}, who analysed 11 bright red giants observed with the WIRE satellite. They compared traditional methods to determine stellar masses with a new method, that uses the effective temperature, the Hipparcos parallaxes and their measurement for \num , to estimate an asteroseismic mass. 
The A2Z pipeline \citep{mat10} uses two different methods to estimate the mass and radius of a star: one based on the scaling laws and the other one that starts with the measurement of \dnu\ and uses a pre-calculated grid of evolutionary models to obtain an initial guess of the fundamental parameters of the star. For the latter method, a minimisation algorithm is performed to estimate the radius and the mass with a higher accuracy \citep{cre07}.
\citet{bas10} presented the Yale-Birmingham (YB) method, which aims to deduce precise stellar radii from a combination of seismic and conventional variables. Within the context of the asteroFLAG hare-and-hounds exercises for the \textit {Kepler Mission}, \citet{ste09b} summarised other methods, which provide stellar radii based on the observed large frequency separation and conventional observables. The basic principle of the YB and asteroFLAG methods is the same. They compare observed seismic parameters (\num\ and/or \dnu ) and other observables (\teff , $V$, $\pi$, \lg , metallicity, etc.) to those of stellar models, where the seismic parameters of the models are determined from scaling relations or adiabatic model frequencies. If the input parameters are well defined, these methods enable very precise estimates for the stellar radius. Most of the red giants observed with CoRoT and \textit{Kepler}, however, are rather faint and, although the seismic parameters can be determined with high precision, additional constraints are generally very uncertain, if available at all. To account for this, \citet{kal10} (hereafter Paper I) presented a modified approach for 31 red giants observed for about 150 days with the CoRoT space telescope. They exclusively used the measured seismic parameters \num\ and \dnu\ to derive estimates for stellar fundamental parameters from the aforementioned scaling relations and a grid of solar-calibrated red-giant models, without making use of any other input parameters. They also indicated that their mass and radius determination is relatively insensitive to the metallicity and/or evolutionary stage of the investigated red giants. 

\begin{figure}[t]
	\begin{center}
	\includegraphics[width=0.5\textwidth]{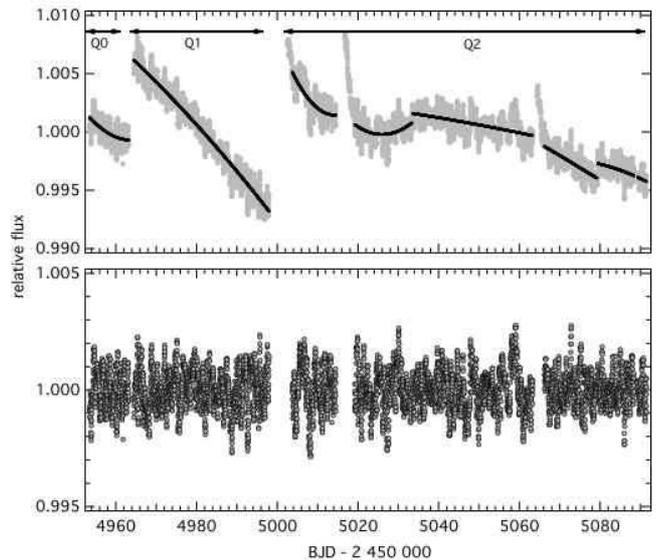}
	\caption{Relative flux for the Q0, Q1, and Q2 data of KIC\,6838420. The top panel shows the original time series (grey points) with the 2nd order polynomial fits overlaid (black lines). The bottom panel shows the residual time series.} 
	\label{fig:lc} 
	\end{center} 
\end{figure}

\begin{figure*}[ht]
	\begin{center}
	\includegraphics[width=1.0\textwidth]{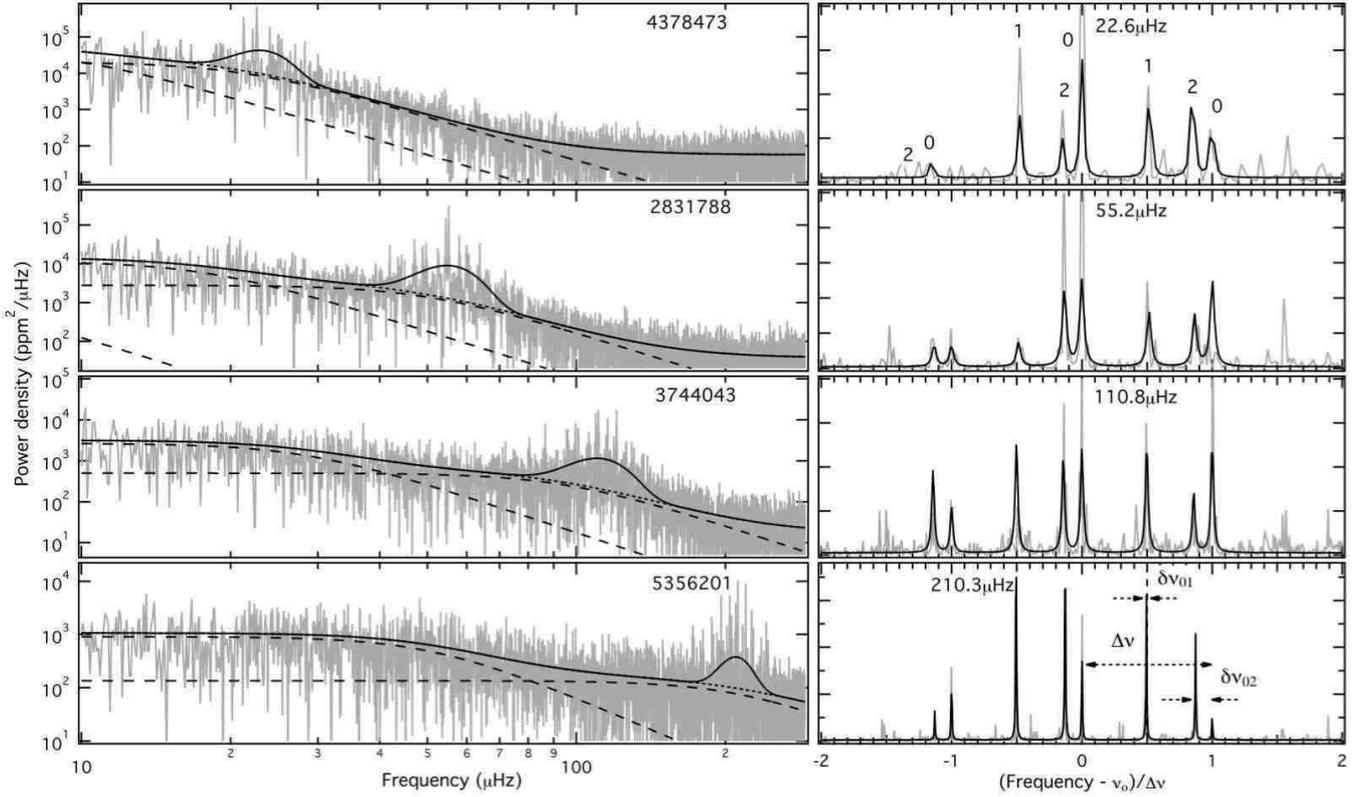}
	\caption{\textit{Left panel}: Power density spectra for a sample of red giants observed with \textit{Kepler}. Black lines indicate the global model fit and dotted lines show the global model plotted without the Gaussian component, which serve as a model for the background signal. Dashed lines indicate the background components. KIC numbers are given in the upper right corners. \textit{Right panel}: Residual power density spectra shifted to the central frequency (given in absolute numbers in the plots) of our model to determine the frequency separations and normalised to the large frequency separation. Black lines correspond to the best fitting model. The dashed line marks the midpoint between adjacent $l$ = 0 modes.} 
	\label{fig:spec} 
	\end{center} 
\end{figure*}

In this paper, we largely follow the approach of Paper I but for red giants that have been observed during the first $\sim$138 days of science operation of the \textit{Kepler} satellite. We fit a global model to the power density spectra of the high-precision photometric time series to measure \num . We use an improved approach to determine the large and small frequency spacings, which also provides an automated identification of the mode degree. We apply the same methods to determine \num\ and \dnu\ to SOHO/VIRGO (\textit{Variability of solar IRradiance and Gravity Oscillations}; \citealt{fro97}) data to measure the solar reference values needed as an input for the scaling relations. We compare the measured seismic parameters with those determined from a multi-metallicity red-giant model grid (by using the scaling relations) to derive a reliable stellar mass and radius and a reasonable effective temperature and luminosity for the red giants in our sample. We investigate the error sources in our analysis and discuss the stellar populations on the giant branch.

\section{Observations}	\label{sec:obs}

The NASA \textit{Kepler Mission} \citep{bor08,bor10} was launched in March 2009 with the primary goal of searching for transits of Earth-sized planets in and near the habitable zones of Sun-like stars. The satellite houses a 95-cm aperture modified Schmidt telescope that points at a single field in the constellation Cygnus for the entire mission lifetime ($>$3.5\,years). The telescope feeds a differential photometer with a wide field of view that continuously monitors the brightnesses of about 150,000 stars. This makes it an ideal instrument for asteroseismology and the Kepler Asteroseismic Science Consortium (KASC)\footnote{http://kepler.asteroseismology.com} has been set up to study many of the observed stars (see \citealt{gil10} for an overview and first results).

In this paper we concentrate on the long-cadence \citep[29.4 minutes sampling;][]{jen10} data that have been collected within the astrometric and asteroseismic programmes during the commissioning phase (Q0; $\sim$11d) and the first (Q1; $\sim$34d) and second (Q2; $\sim$89d) roll of the satellite. We analyse all stars for which at least Q1 and Q2 time series have been made available via KASC, including those which have not been flagged as red giants in the Kepler Input Catalogue \citep[KIC;][]{lat05}. The combined time series consist of about 5900 or 5430 measurements and span a total duration of about 138 or 127 days, depending on the availability of Q0 data. 

In Fig.\,\ref{fig:lc} we show the relative flux for a typical red giant. Whereas the Q0 and Q1 time series show only long term trends, the Q2 data reveal a more complex behaviour. The time series appears to consist of five parts with sudden jumps at the transitions, which are instrumental. Additionally there is a step gradient at least at the beginning of the second and fourth ``subset'' (BJD - 2\,450\,000 $\simeq$ 5018 and 5062). These artefacts appear at the same time in most data sets and are due to some satellite failures during the Q2 observations, where \textit{Kepler} had to be cooled down again after safe-mode operations. To account for these instrumental artefacts, we split the time series into 7 subsets (Q0, Q1, and 5 subsets for Q2). We tried several approaches to model the gradients, but this turned out to be quite difficult as the actual shape differed from star to star, even increasing gradients have been found in some cases. Therefore, we simply removed the leading data points, including the steepest part of the gradient for the first, second, and fourth subsets of the Q2 data. In total we rejected about 5.5 days of measurements, degrading the overall duty cycle from about 91 to 87\%, with only minor consequences on the spectral window function. Finally, we subtracted a second-order polynomial fit from each subset. The resulting time series is shown in the bottom panel of Fig.\,\ref{fig:lc}. This approach does, of course, suppress any intrinsic long period signal. The shortest subset is about 11 days long, which means that we filter out signal below about 1\mh .

The long-cadence data from \textit{Kepler} that are accessible to KASC consist of two major samples. Firstly, the so-called astrometric reference stars \citep{bat10,mon10} comprising about 1\,000 stars that have been selected to be distant (and therefore having a small parallax), but bright (and therefore being mostly giants) and unsaturated stars in a \textit{Kepler} magnitude range of 11.0--12.5\,mag, which are uncrowded and uniformly distributed over the focal plane. Secondly, about 1\,300 stars that have been selected for asteroseismology by the various working groups of KASC according to different criteria such as their presumed membership to a cluster or due to their colour index. We computed the power density spectra between 1\mh\ and the Nyquist frequency ($\sim$280\mh ) for all stars and searched them visually (i.e., by eye) for red-giant characteristics. We found a total of 1041 stars (670 astrometric and 371 asteroseismic) that show both a clear power excess hump with regularly spaced peaks and a background that decreases towards higher frequencies. We identified them as red giants for the subsequent analysis.

\section{Power spectra modelling}	\label{sec:powspec}

The power spectra of solar-type oscillations have characteristic features. Besides an instrumental white noise component, they show a frequency-dependent background signal. This signal can be represented by several super-Lorentzian functions\footnote{Note that this function is frequently referred to as power law or Harvey-like model. It is, however, clearly not a power law and \citet{har85} originally used a Lorentzian. We therefore suggest the name ``super-Lorentzian" with the power 4, which is sometimes used in optics.}
with increasing characteristic frequencies and decreasing characteristic amplitudes. Each of these components is believed to represent a separate class of physical process such as stellar activity and the different scales of granulation, and most of them are strongly connected to the turbulent motions in the convective envelope. On top of the background signal, one finds additional power due to pulsation in a broad hump. This power excess arises from a sequence of stochastically excited and damped oscillations, which correspond to high-overtone radial and non-radial acoustic modes. The mode amplitudes are defined by the excitation and damping and are modulated by a broad envelope. The centre of the envelope is usually called the frequency of maximum oscillation power (\num ) and its shape is approximately Gaussian. 

\subsection{Global power spectrum model and \num }

The background signal in the power spectra of solar-type oscillations can be modelled by the sum of super-Lorentzian functions, $P(\nu) = \sum A_i / (1 + (2\pi \nu \tau_i)^{c_i})$, with $\nu$ being the frequency, $A_i$, $\tau_i$, and $c_i$ being the characteristic amplitudes, timescales, and the slopes of the background model. This model was first introduced by \citet{har85} to characterise the solar background signal. In Paper I, it was shown that the solar background model also works for the power spectra of red giants. Due to the larger radii of red giants compared to the Sun, the amplitudes and timescales of the background components are quite different but the model, particularly the slopes of the components are the same. Here, we follow the approach of Paper I and model the observed power density spectra with a superposition of white noise, the sum of super-Lorentzian functions, and a power excess hump approximated by a Gaussian: 

\begin{equation}
P(\nu) =  P_\mathrm{n} +  \sum_{i} \frac{2\pi a_{i}^2/b_i}{1 + (\nu / b_i)^4} + P_\mathrm{g}  \exp{\left ( \frac{-(\nu_\mathrm{max} - \nu)^2}{2\sigma_g^2} \right )}
\label{eq:bg}
\end{equation}
where $P_\mathrm{n}$ corresponds to the white noise contributions and $a_i$ is the rms amplitude of the $i$th background components. The parameter $b_i$ corresponds to the frequency at which the power of the component is equal to half its value at zero frequency and is called the characteristic frequency. $P_\mathrm{g}$, \num , and $\sigma_g$ are the height, the central frequency, and the width of the power excess hump, respectively. Note that $\sigma_g$ is about 1.18 times the HWHM. For our sample of red giants, the frequency coverage of the \textit{Kepler} observations allowed us to model up to three background components.

The only difference compared to the model in Paper I is the numerator in the background models. Originally, a single parameter, $A$, was used which corresponds to the power at frequency equal to zero of the given component. However, tests have shown that fitting $a$ and $b$ instead of fitting $A$ and $b$, with $A = 2\pi a^2/b$, yields a more robust fit and allows a more accurate measurement of the characteristic frequencies. This notation also makes more sense physically because $a^2$ corresponds to the variance that the signal produces in the time domain and can easily be related to the observed total energy of, e.g., granulation.  
\begin{figure}[b]
	\begin{center}
	\includegraphics[width=0.5\textwidth]{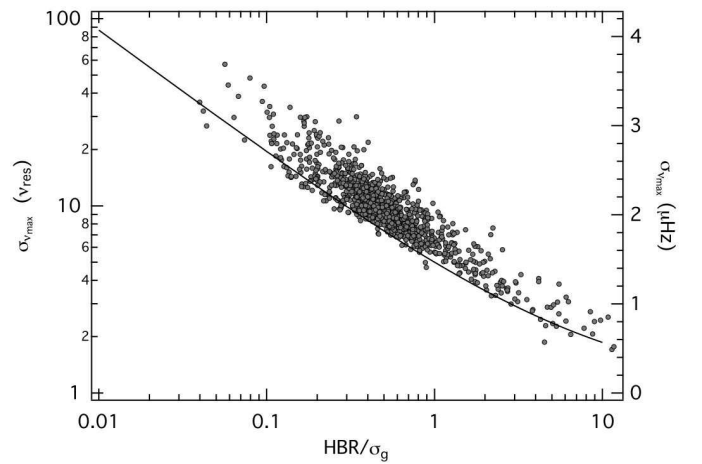}
	\caption{The uncertainty in \num\ in actual value (right axis) and in units of the frequency resolution as a function of the ratio between the height-to-background ratio (HBR) and the width of the power excess, $\sigma_g$. The line indicates a power-law model for the lower envelope.} 
	\label{fig:numaxerror} 
	\end{center} 
\end{figure}

We used a Bayesian Markov-Chain Monte-Carlo (MCMC) algorithm to fit the global model to the power density spectra. See Paper I and \citet{gru09} for a detailed description. Briefly, the algorithm automatically samples a wide parameter space and delivers probability density distributions for all fitted parameters and their marginal distributions, from which we computed the most probable values and their 1$\sigma$ uncertainties. For the parameter limits, we followed a slightly modified approach to that in Paper I. During the fitting process we kept \num\ within $\pm$25\% of the value inferred from the visual inspection of the spectrum. The width of the power excess was allowed to vary between 5\% and 50\% of the initial guess of \num , where the lower limit prevented the algorithm from fitting the Gaussian to a single frequency bin in the spectrum. The frequency parameters $b_i$ were allowed to vary from 0 to 1.5 times \num , with the condition $b_1 > b_2 > b_3$, where the indices indicate consecutive background components. The amplitude parameters $a_i$ were kept between 0 and 10 times the square root of the highest peak power in the spectra. $P_g$ was allowed to vary from zero to 10 times the average power in the spectrum around the initial guess for \num , and $P_n$ was kept between 0.5 and 2 times the average power at the high frequency end of the spectrum. The left panel of Fig.\,\ref{fig:spec} shows examples of power density spectra with the corresponding fits. 
\begin{figure}[t]
	\begin{center}
	\includegraphics[width=0.5\textwidth]{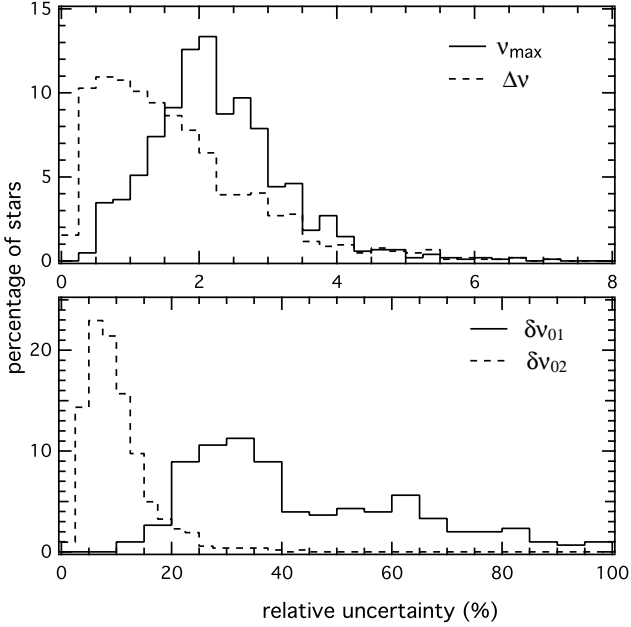}
	\caption{Histograms of the relative uncertainties for \num , the large frequency separation \dnu , and the small separations $\delta\nu_{01}$ and $\delta\nu_{02}$.} 
	\label{fig:numaxerrorhisto} 
	\end{center} 
\end{figure}

An important aspect of our analysis is to understand the uncertainties of the determined parameters. One might expect that the white noise, and therefore the brightness of a star, is responsible for a significant part of the uncertainty in \num\  but we do not find any correlation with the magnitude. 
On the other hand, there is a clear correlation between $\sigma_{\nu\mathrm{max}}$ and the ratio of the height-to-background ratio (HBR) to the width of the power excess ($\sigma_g$), where we define the HBR as the ratio between the height of the power excess ($P_g$) and the background signal at \num . In other words, we can more accurately determine the centre of a narrow power excess hump with a large HBR than the centre of a broad hump with a small HBR. This is illustrated in Fig.\,\ref{fig:numaxerror} where we plot the absolute value of $\sigma_{\nu\mathrm{max}}$ (right axis) as a function of HBR/$\sigma_g$. Tests with subsets of the \textit{Kepler} time series (and data sets from CoRoT) have shown that $\sigma_{\nu\mathrm{max}}$ is also directly proportional to the frequency resolution of the time series, which is defined as the inverse data set length. To account for this we plot $\sigma_{\nu\mathrm{max}}$ in units of the frequency resolution ($\nu_\mathrm{res}$) on the left axis in Fig.\,\ref{fig:numaxerror}. From that, we can define a simple relation for the lower limit of the uncertainty in \num\ as 
\begin{equation}
\sigma_{\nu\mathrm{max}} = \nu_\mathrm{res}  \left (1 + \frac{4}{(HBR/\sigma_g)^{2/3}} \right ). 
\end{equation}
For the solar case we would expect an uncertainty in \num\ of about 5.3\mh , which is in good agreement with the value of 5.23\mh\ found in Sect.\,\ref{sec:sun} for SOHO/VIRGO data. Given this relatively simple error law we are confident that our uncertainties for \num\ are reliable and do mainly reflect the constraints of the observations and not of our method. It is also interesting to see that the uncertainty of \num\ is very much defined by the star itself, since HBR/$\sigma_g$ is largely intrinsic to the star, for a given instrument and observing time. Nevertheless, we mention that the error law is purely phenomenological and might not be valid outside the range we use it.

A histogram of the relative uncertainties in \num\ is given in the top panel of Fig.\,\ref{fig:numaxerrorhisto}, showing a clear peak at about 2\%. This is not surprising as a large fraction of the analysed red giants are red-clump stars having a very similar \num , and therefore a similar width and height-to-background ratio of the power excess. We see that, for almost all stars, we could determine \num\ to within 4\% and, for about half of our sample, to within 2\%.

A potential problem for the subsequent analysis is that we assumed the power excess hump to be symmetric, so that an intrinsic asymmetry might result in a systematic error. In Paper I it was claimed that the asymmetry of the power excess hump is within the observational uncertainties of \num , and therefore negligible. However, that conclusion was based on the analysis of only 31 stars. Our sample is more than 30 times larger and should give a statistically more significant conclusion. We computed the weighted mean frequency, $\nu_\mathrm{wm}$, in the frequency range of pulsation (\num $\pm$3$\sigma_g$), where we used the residual power after correcting for the background signal as weight. We found $\nu_\mathrm{wm}$ consistently shifted towards higher frequencies compared to \num\ by 3.1$\pm$1.3\% and therefore outside the average uncertainty for \num\ of about 2.3\%. This is, however, not a problem in our subsequent analysis since we find the same shift of about 3\% in the solar data (see Sect.\,\ref{sec:sun}), and as long as we compare \num\ values that are determined in the same way, we do not have to take into account asymmetries in the power excess humps.

\subsection{Frequency spacings} \label{sec:spacings}

In the next step we used the white noise and background components of the global model (dotted lines in Fig.\,\ref{fig:spec}) to correct the power density spectra for the background signal, leaving only the oscillation signal and white noise. The second parameter that can directly be determined from the observed power spectrum is the large frequency separation, \dnu . To determine \dnu , we use a similar approach as in Paper I and fit the following general model to the residual power density spectrum over a frequency range spanning three radial orders around the frequency of maximum oscillation power:  
\begin{equation}\label{eq:spacings}
\begin{split}
P(\nu) =  P_n & +  \sum_{i=-1}^{1} \frac{A_i^2  \tau}{1+ 4  [\nu - (\nu_0 + i \Delta\nu)]^2  (\pi\tau)^2} \\
                        & +  \sum_{j=-1}^{1} \frac{A_j^2  \tau}{1+ 4  [\nu - (\nu_0 + j \Delta\nu - \delta\nu_{02})]^2  (\pi\tau)^2}\\
                        & +  \sum_{k=-1,1}   \frac{A_k^2  \tau}{1+ 4  [\nu - (\nu_0 + \frac{k}{2} \Delta\nu + \delta\nu_{01})]^2  (\pi\tau)^2}.\\
\end{split}
\end{equation}
The model represents a sequence of eight Lorentzian profiles whose frequencies are parameterised by a central frequency, $\nu_0$, and three spacings, \dnu , $\delta\nu_{01}$, and $\delta\nu_{02}$, where the first, second, and third sum corresponds to three radial, three $l$ = 2, and two $l$ = 1 modes, respectively. $A_{i}$, $A_j$, and $A_k$ are the individual rms amplitudes. As in Paper I, we assume the same mode lifetime $\tau$ for all modes, which might not be true in reality, but this assumption has no impact on the determination of the spacings and significantly stabilises the fit. We again used the Bayesian MCMC algorithm to fit the model to the residual power density spectrum. All mode amplitudes were allowed to vary independently between zero and 10 times the highest peak in the amplitude spectrum. This allowed the algorithm to account for missing modes or modes hidden in the noise. The mode lifetime was sampled between 1 and 100 days. Most important here were the parameter ranges for the spacings. With the condition that $\delta\nu_{01} \ll $ \dnu\ and $\delta\nu_{02} <$ \dnu /2, the model basically represents the asymptotic relation \citep{tas80} for low-degree and high-radial order p modes. Consequently, $\delta\nu_{02}$ and $\delta\nu_{01}$ correspond to the small separations between adjacent $l$ = 0 and 2 modes and between $l$ = 1 modes and the midpoint of consecutive radial modes, respectively. A critical parameter is the central frequency because the central mode must be a radial mode. Otherwise, the interpretation of the spacings contradicts the asymptotic relation. It turned out when allowing $\nu_o$ and \dnu\ to vary over a relatively wide range (\num\ $\pm$2$\sigma_g$ for $\nu_o$ and 0.5\mh\ to 2$\sigma_g$ for \dnu ) the algorithm was able to automatically find the central mode that corresponds to a radial mode if $l$ = 2 modes were present. In case of no detectable $l$ = 2 modes, the mode identification is ambiguous, but the delivered large spacing was still a good estimate. 

We have determined \dnu\ for all 1041 stars and found at least two $l$ = 2 modes in the oscillation spectrum of about half of our sample, which allowed us to determine small spacings based on an automated mode identification. In principle, we have also $\delta\nu_{01}$ values for these stars but only accepted them for about one third of the total sample. This is because of the multiple dipole modes over a relatively broad frequency range per order, which occur due to their mixed gravity/acoustic mode character \citep{dup09}, and makes it difficult for our algorithm to obtain robust results. We do not further investigate the small spacings here but refer to \citet{hub10}, where those results are presented. 

The oscillation spectra and the corresponding best fits for the stars in Fig.\,\ref{fig:spec} are illustrated in the right panel of that figure. As for our global fit parameters, we determined the most probable parameters from the marginal distributions of the probability density delivered by the MCMC algorithm. Unlike for \num , we were not able to find a clear correlation between the uncertainty in \dnu\ and any other parameter combination. We expect, however, that $\sigma_{\Delta\nu}$ depends on the frequency resolution, the signal-to-noise ratio of the individual modes and their lifetime. We could not reliably determine the mode lifetimes for many stars because their mode profiles are undersampled, which means that the peak width due to the spectral window function of the observations is broader than the actual profile width. In such a case our amplitudes and lifetimes are meaningless. The mode frequencies and therefore the spacings are, however, not affected by this phenomenon. Histograms for the uncertainties of the spacing parameters are given in Fig.\,\ref{fig:numaxerrorhisto}, showing that we can determine \dnu\ to within 1\% for about 30\% of our sample. Whereas the accuracy of $\delta\nu_{02}$ is mostly better than 10\%, the relative errors for $\delta\nu_{01}$ are relatively large. But one has to keep in mind that $\delta\nu_{01}$ is well below 1\mh\ for a red giant and the absolute uncertainties of $\delta\nu_{01}$ are still quite low.

Originally, we would have had to exclude a number of stars from our sample because they pulsate with low frequencies making it difficult to reliably determine spacings from a frequency spectrum. There is, however, the so-called autocorrelation method described by \citet{mos09}, which measures \dnu\ from the first peak in the autocorrelation of the time series. This method is less affected by the limited duration of the observations than Fourier-based methods and is able to detect regular spacings down to a few times the frequency resolution \citep{mos10}. 
To account for this, we used the \dnu\ values from the autocorrelation method for all stars with \num $<$ 15 \mh\ ($\sim$7\% of the total sample) for the subsequent analysis.

Finally, we cross-checked our results for \num\ and \dnu\ with those of other methods (\citealt{hek10c}, \citealt{hub09}, \citealt{mat10}, and \citealt{mos09}) which have been used to determine the same seismic parameters for our sample (or subsample) of red giants. A direct comparison of the different methods shows that there are a number of outliers but for most stars in our sample, at least one other method gave a value that is compatible with our results (i.e., within the uncertainties). Additionally, we have carefully checked by hand the reliability of the seismic parameters for all stars for which we found a significant disagreement in the direct comparison (see \citealt{hek10b} for the detailed comparison) and identified only a few stars (less than 1\%) which we had to eliminate from our sample because their seismic parameters are ambiguous.


\begin{figure}[t]
	\begin{center}
	\includegraphics[width=0.5\textwidth]{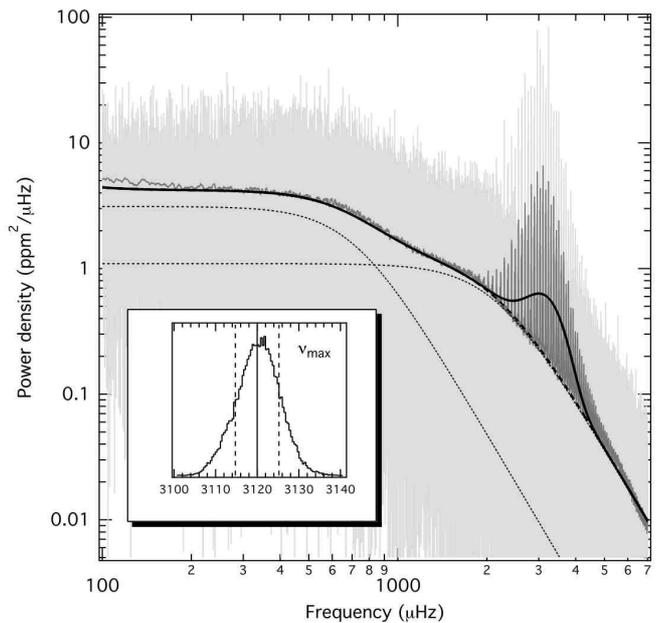}
	\caption{Power density spectrum of a 1-year VIRGO time series (light-grey) and the corresponding global model fit (black line). The dark-grey line shows a smoothed (5\mh\ boxcar filter) version from the average of nine consecutive 1-year VIRGO time series. The inset gives the probability density function for \num\ with the vertical solid and dashed lines indicating the median value and the 1$\sigma$ limits, respectively.} 
	\label{fig:solar1} 
	\end{center} 
\end{figure}

\begin{figure*}[t]
	\begin{center}
	\includegraphics[width=0.95\textwidth]{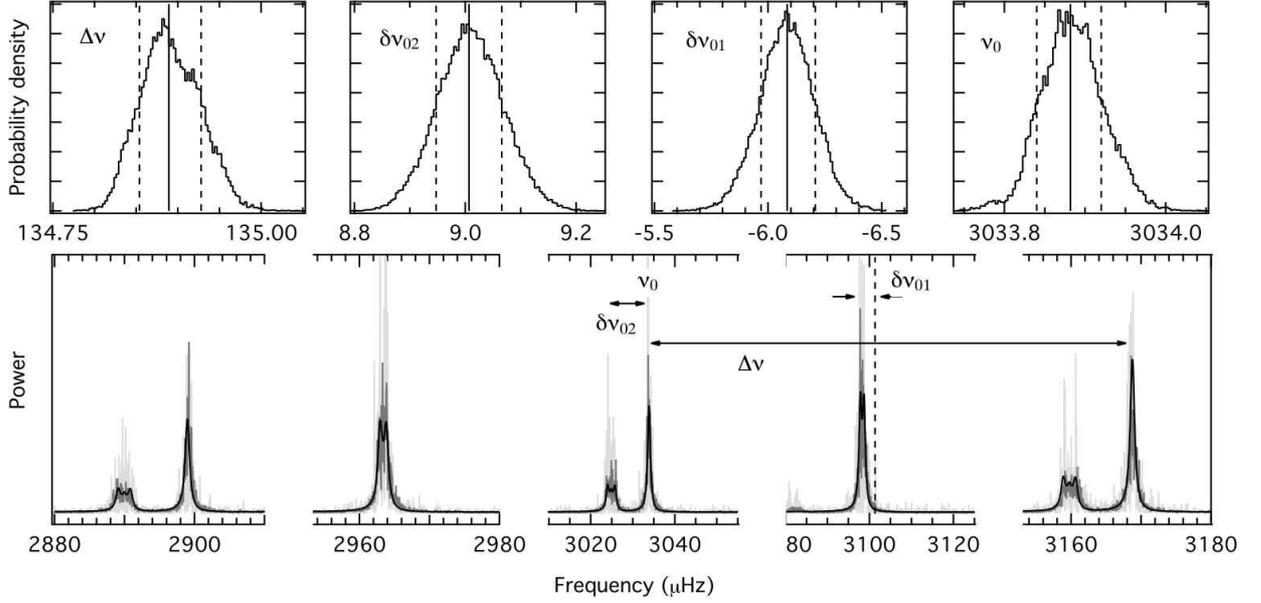}
	\caption{Same as Fig.\,\ref{fig:solar1} but as delivered from our algorithm to determine the frequency spacings. \textit{Top panels}: Probability density functions for the three spacing parameters and the central frequency of our model fit. Median values and 1$\sigma$ limits are indicated by vertical solid and dashed line, respectively. \textit{Bottom panel}: Residual power density spectrum (light-grey) of a 1-year VIRGO time series and the corresponding model fit (black line).} 
	\label{fig:solar2} 
	\end{center} 
\end{figure*}

\subsection{Solar reference parameters} \label{sec:sun}

The large frequency separation is related to the inverse sound travel time through the star and therefore to the mean density of the star.
It scales as \dnu $_{\sun} (M/R^3)^{1/2}$ from the solar case, with $R$ and $M$ being the total mass and the radius of the star, respectively, in solar units. An important point when using scaling relations to estimate fundamental parameters is the definitions of the scaled seismic parameters. An often used solar reference value, \dnu = 134.95\mh\ is based on the frequency difference between the radial modes with order n = 20 and 21 where the maximum oscillation power is seen \citep{tou92}. The frequency difference of two single modes is difficult to determine for other stars and an average value of all (or some) observable modes is often used (e.g, $\sim$134.8\mh\ for the Sun; \citealt{kje08}). However, the frequency separation is a function of the frequency itself (see e.g. \citealt{bro09} for the Sun and \citealt{mos10} for CoRoT red giants), and an average \dnu\ will depend on the actual number and frequency range of the observed modes and is difficult to compare for different observations.

The frequency of maximum oscillation power, on the other hand, is related to the acoustic cut-off frequency in the stellar atmosphere \citep[e.g.,][]{bro91,kje95}, which in turn scales as \num $_{,\sun} (M\,\,R^{-2}\,\,T${\tiny eff}$ ^{-1/2})$ from the solar case, where often used values are \num $_{,\sun}$ = 3050\mh\ \citep[e.g.,][]{kje95} or 3100\mh\ \citep[e.g.,][]{bas10}. Both seismic parameters scale with the stellar mass and radius and can therefore be used to estimate $R$ and $M$ of a star from its seismic parameters. But to use these seismic scaling relations in a consistent way we need values for the solar parameters, measured in the same way as for our sample of red giants.  
Note that the seismic scaling relations are not laws of physics and possibly include some additional uncertainties. There are, however, strong indications that at least the \dnu -scaling is quite accurate for cool stars like our sample of red giants \citep{bas10, ste09c}.
	
We used a 1-year time series from the green channel of the SOHO/VIRGO data \citep{fro97} obtained during a solar activity minimum and fitted our global model (Eq.\,\ref{eq:bg}) to the corresponding power density spectrum. The result is given in Fig.\,\ref{fig:solar1}, showing the original power density spectrum and a smoothed version of the average power density spectrum of nine consecutive 1-year time series along with the best fitting model. Although the model is fitted to ``only" a 1-year time series, it almost perfectly reproduces the average spectrum of the full 9-year time series. The reason why we do not use the 9-year time series is because of limited computer resources. Also shown is the probability density function of the solar \num\ parameter, from which we determined \num $_{,\sun}$ = 3120$\pm$5\mh .
	
In the next step we use the background part of the global model fit to correct the power density spectrum and fit the asymptotic relation model to the residual spectrum. The solar p-mode profiles are naturally much better resolved than the p-modes in our sample of red giants and initial tests have shown that the rotationally split components of the non-radial modes significantly disturb the fit. To account for this we have included rotationally split components for the $l$ = 1 and 2 modes, which are parameterised by a single rotational frequency and an inclination angle that defines the height ratio between the central profile and the split components \citep[e.g.,][]{giz03,bal06}. We have also tested asymmetric mode profiles \citep{nig98} but found no significant influence on the parameters of interest. In Fig.\,\ref{fig:solar2}, we show the probability density distributions for the three frequency spacings  and the central frequency of our model, together with their median values and 1$\sigma$ limits. The bottom panels illustrate the residual power density spectrum and the best model fit. We have determined \dnu $_{\sun}$ = 134.88$\pm$0.04\mh , $\delta\nu_{02,\sun}$ = 9.00$\pm$0.06\mh , and $\delta\nu_{01,\sun}$ = 6.14$\pm$0.10\mh . The uncertainties might appear unrealistically small but they reflect a well defined case specifically chosen for our approach and should not be mistaken as ``global" frequency spacings of the Sun.

\section{Asteroseismic fundamental parameters} 	\label{sec:fund}
\subsection{Method}
Our approach to determining an asteroseismic mass and radius depends on the aforementioned scaling relations for \num\ and \dnu  :
\begin{align} 
 \nu_\mathrm{max} & = 3120\pm5\mu \mathrm{Hz} \times M R^{-2} T^{-1/2}  \\
 \Delta\nu & = (134.88\pm0.04\mu \mathrm{Hz}) \times M^{1/2}R^{-3/2}  
\end{align}
with $M$ and $R$ in solar units and $T = T_\mathrm{eff} /  5777\,K$. Obviously the \num\ scaling relation depends on the effective temperature of the star. Although there are temperatures available in the KIC for most of the analysed stars, we decided not to directly use them in our analysis (i.e., in the above scaling relations). The KIC parameters (\teff , \lg , and [Fe/H]) are mostly determined from multi-colour photometry and they are calibrated to be correct in a statistical sense, i.e., the average values of a large sample of stars are correct. However, a star-by-star comparison (e.g., for stars associated in a cluster) has shown that the individual values can have a large scatter (up to some 100\,K in \teff ). Consequently, adopting the KIC temperatures would add uncertainties of up to 5\% and 10\% to our radius and mass determinations, respectively. Therefore, we also compared the seismic parameters of the observed stars and their KIC temperatures to those inferred from stellar models, and then adopted the fundamental parameters from the model that best reproduced all input parameters. There are several ways to do this. In Paper I, an initial guess for the stellar mass and radius was determined using an average temperature for stars on the red giant branch. The initial mass and radius values were then compared to those of a grid of solar-calibrated red-giant models to get a better estimate for the temperature. The new temperature was adopted and the procedure repeated, converging to a certain location in the H-R diagram after a few iterations and delivering a full set of fundamental parameters for each star. However, these iteratively determined parameters also depend on the chemical composition and the evolutionary status of the model (i.e., whether the model is a red-giant branch or an asymptotic-giant branch model). This ambiguity allows a given set of seismic parameters to converge to different locations in the H-R diagram if different model grids with, e.g., a different chemical composition are used. 
This uncertainty has, however, only small effects on the mass and radius determination. For the effective temperature and luminosity it adds about $\pm$150\,K and $\pm$20\% uncertainties, respectively.

A more general approach was presented by \citet{bas10}. They used a combination of seismic and conventional stellar parameters and compared them to those of a multi-metallicity model grid, where the aforementioned scaling relations are used to determine the seismic parameters for the models. \citet{bas10} defined a model likelihood from the difference between the model and input parameters and infer the stellar radius from the resulting likelihood function. From their tests with a number of artificial stars they concluded that it is very unlikely to get a reasonable estimate for a red giants' radius if no accurate temperature and parallax are available. This conclusion was, however, based on only a single red-giant test star that is located high up the red giant branch and therefore quite far away from where most of the observed red giants are expected (i.e. in the red clump). Additionally, they have assumed uncertainties for the seismic input parameters that are about ten times larger than what we have determined for our sample of red giants.
\begin{figure}[b]
	\begin{center}
	\includegraphics[width=0.5\textwidth]{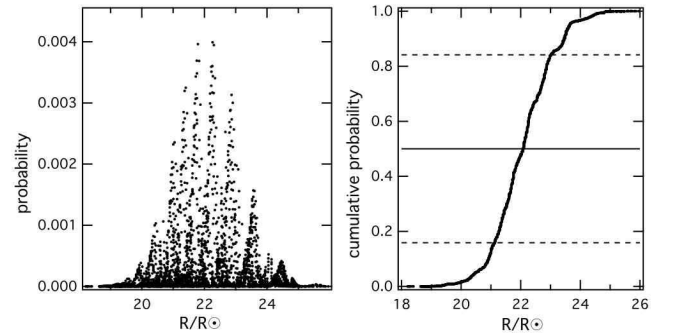}
	\caption{The projected probability (left) and cumulative probability (right) distribution functions for the radius of the artificial test star 6 \citep{bas10}. The full horizontal line correspond to the median and the dashed lines give the $\pm$1$\sigma$ confidence interval.} 
	\label{fig:teststar} 
	\end{center} 
\end{figure}

Here, we follow the approach of \citet{bas10} but formulate our search for a best set of fundamental parameters in a Bayesian sense. Given a set of seismic input parameters $\nu_\mathrm{max, obs}$ and $\Delta\nu_\mathrm{obs}$, and the Gaussian distribution of their uncertainties, $\sigma_{\nu\mathrm{max}}$ and $\sigma_{\Delta\nu}$, we define the likelihood that the seismic model parameters, $\nu_\mathrm{max, model}$ and $\Delta\nu_\mathrm{model}$, matches the observed ones as
\begin{align}\label{eq:likelyhood}
\mathcal{L}_{\nu_\mathrm{max}} & = \frac{1}{\sqrt{2\pi}\sigma_{\nu_\mathrm{max}}}  \exp{ \left ( \frac{-(\nu_\mathrm{max,obs} - \nu_\mathrm{max,model})^2}{2\sigma_{\nu_\mathrm{max}}^2} \right)}\\
\mathcal{L}_{\Delta\nu} & = \frac{1}{\sqrt{2\pi}\sigma_{\Delta\nu}}  \exp{ \left (\frac{-(\Delta\nu_\mathrm{obs} - \Delta\nu_\mathrm{model})^2}{2\sigma_{\Delta\nu}^2} \right)}.
\end{align}
To simplify matters, we also assume a Gaussian error for the KIC effective temperatures and define the likelihood that the model matches the observed temperature as
\begin{equation}
\mathcal{L}_{T_\mathrm{eff}}  = \frac{1}{\sqrt{2\pi}\sigma_{T_\mathrm{eff}}}  \exp{ \left (\frac{-(T_\mathrm{eff,KIC} - T_\mathrm{eff,model})^2}{2\sigma_{T_\mathrm{eff}}^2} \right)},
\end{equation}
with a typical value of 200\,K for $\sigma_{T\mathrm{eff}}$.
In the Bayesian approach we can assign an overall probability of the model $M_i$ with the posterior probability $I$ matching the observed parameters $D$ with respect to the entire set of models according to BayesÕ theorem as
\begin{equation}\label{eq:bayes}
p(M_i|D,I) = \frac{p(M_i|I) p(D|M_i,I)}{p(D|I)}
\end{equation}
where 
\begin{equation}
p(M_i|I) = \frac{1}{N_m}
\end{equation}
is the uniform prior probability for a specific model with $N_m$ being the total number of models, and
\begin{equation}
p(D|M_i,I) = \mathcal{L}(\nu_\mathrm{max},\Delta\nu, T_\mathrm{eff}) = \mathcal{L}_{\nu_\mathrm{max}} \mathcal{L}_{\Delta\nu} \mathcal{L}_{T_\mathrm{eff}}
\end{equation}
is the likelihood function. The denominator of Eq.\,\ref{eq:bayes} is a normalisation factor for the specific model probability in the form of
\begin{equation}
p(D|I) = \sum_{j=1}^{N_m} p(M_j|I) \cdot p(D|M_j,I).
\end{equation}
Since the uniform priors are the same for all models they cancel in Eq.\,\ref{eq:bayes}, which simplifies to 
\begin{equation}
p(M_i|D,I) = \frac{p(D|M_i,I)}{\sum_{j=1}^{N_m} p(D|M_j,I)}.
\end{equation}
The resulting model probability distribution automatically translates into most probable fundamental parameters and their uncertainties by constructing the marginal distribution for the corresponding parameter. The normalised probability of the most probable parameters is therefore a measure of how likely they are with respect to the other models of the specific grid. We stress that it does not tell us how probable the parameters are in an absolute sense, although formally it must be implicitly assumed than one of the tested models is actually true. The probability must be interpreted as being restricted to the space of the models being considered, and their associated physics. This means that as soon as additional models are added to the model space, the probabilities will change. 

\begin{figure}[t]
	\begin{center}
	\includegraphics[width=0.5\textwidth]{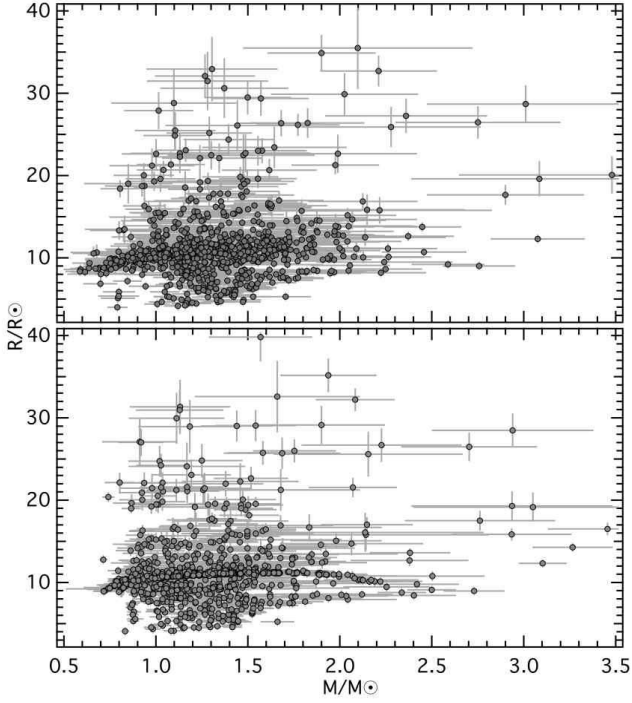}
	\caption{The stellar radius as a function of mass for the sample of red giants as directly determined from Eq.\, 4 and 5 using the KIC effective temperatures (top) and from the comparison with the stellar model grid (bottom). } 
	\label{fig:RMcomp} 
	\end{center} 
\end{figure}
We also mention for completeness that since we use an uniform prior that rates each model with the same prior probability our Bayesian approach is actually not very different from the likelihood method of \citet{bas10}. But as soon as we would add additional informations such as, e.g., an initial mass function, the advantages of the Bayesian technique would become obvious. But this is beyond the scope of the present paper and we leave it to future investigations.
\begin{figure}[h]
	\begin{center}
	\includegraphics[width=0.47\textwidth]{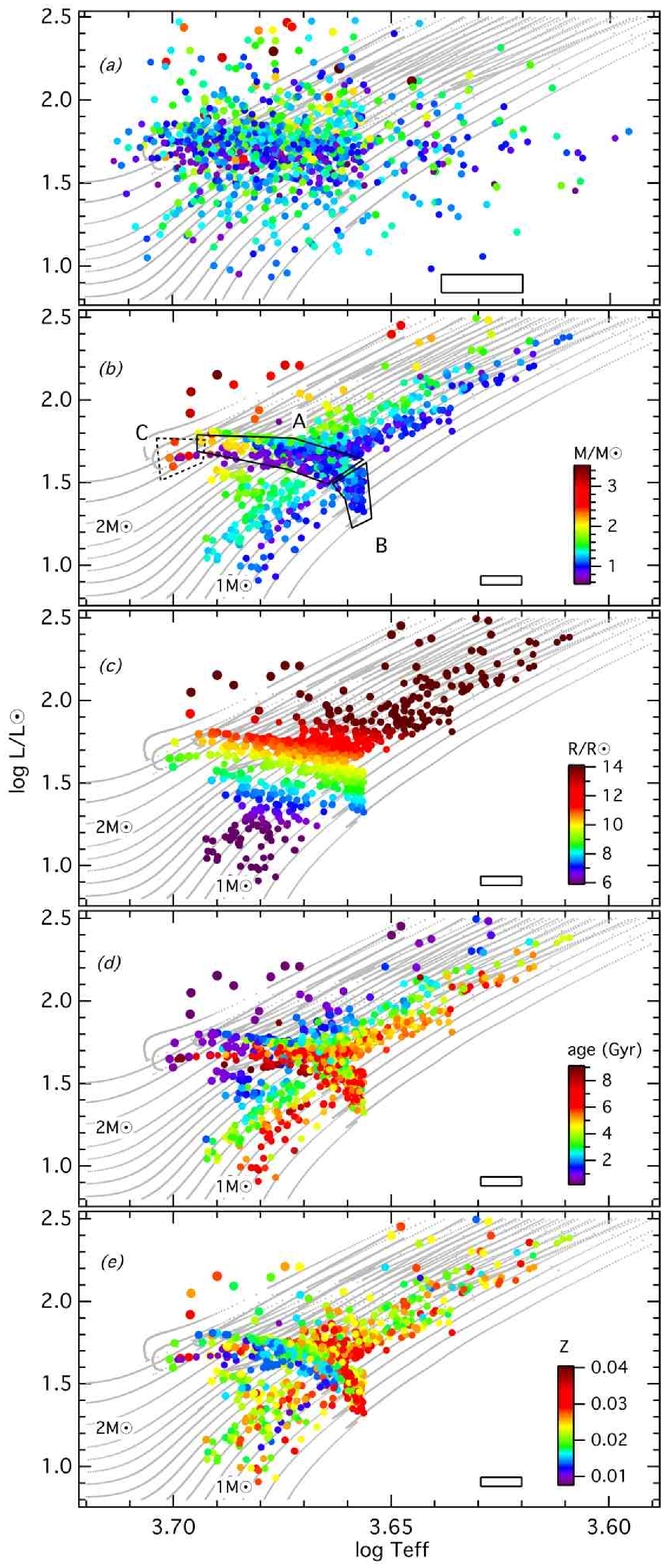}
	\caption{H-R diagram showing the location of our red-giant sample as directly determined from Eq.\, 4 and 5 using the KIC effective temperatures (panel $a$) and from the comparison with the stellar model grid (panel $b$ to $e$). The colour code (in the online version only) indicates the stellar mass ($a$ and $b$), radius ($c$), age ($d$), and metallicity ($e$), where the radius scale has been truncated below 6 and above 14\,R\sun . Grey lines show solar-metallicity BaSTI evolutionary tracks. The boxes given in the lower right corners illustrate typical uncertainties.} 
	\label{fig:HRDcomp} 
	\end{center} 
\end{figure}

\subsection{Models}
The red-giant models used for our analysis were extracted from the canonical scaled-solar BaSTI\footnote{Available from http://albione.oa-teramo.inaf.it/.} isochrones \citep{pie04} in version 5.0.0 with a mass-loss parameter $\eta$ = 0.2. The grid includes models which were evolved from the zero-age main-sequence up to the tip of the red-giant branch (RGB), down to the He-core burning main sequence (red clump) and back up to the asymptotic giant branch (ABG) to an age of about 15\,Gyr. We restricted the grid to models that have already left the main sequence with initial chemical compositions of (Z, Y) = (0.008, 0.256), (0.01, 0.259), (0.0198, 0.2734), (0.03, 0.288), and (0.04, 0.303). The model mass ranges from 0.7 to 4\,M\sun\ with steps of typically 0.02 \,M\sun . We rejected models both with a mass above 4\,M\sun\ and with metallicities below Z = 0.008 because they develop a distinctive horizontal giant branch that would significantly complicate our analysis. 
\begin{figure*}[t]
	\begin{center}
	\includegraphics[width=1.0\textwidth]{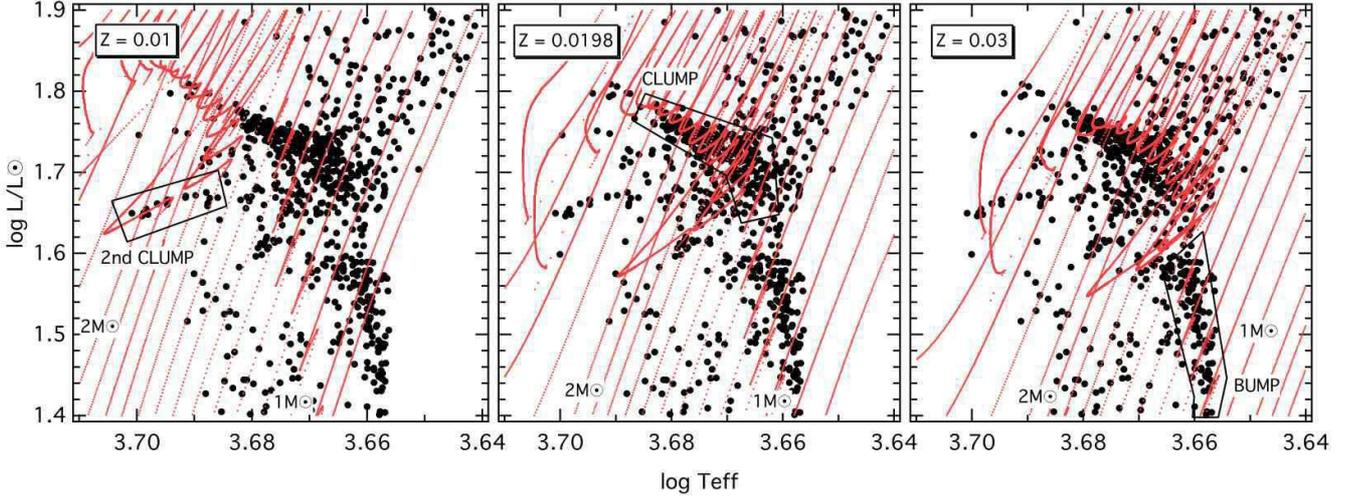}
	\caption{H-R diagrams showing our sample of red giants (black dots) with respect to different metallicity model grids (red dots - in the online version only).} 
	\label{fig:HRD_Z} 
	\end{center} 
\end{figure*}
We think this is justified because very high mass and very metal-poor stars are both quite rare and it is unlikely to find more than a few in a random sample of red giants. Initial tests showed that the poor resolution in chemical composition leads to an artificial clustering of stars in the H-R diagram, i. e., the red-clump stars are concentrated in sharp features corresponding to the red-clump models of the different metallicity grids. We therefore increased the resolution in chemical composition to steps of 0.002 in Z via interpolating the effective temperature for models of approximately constant mass and radius but with different chemical composition. The corresponding luminosities were determined from the Stefan-Boltzmann law, $L \propto R^2 T^4$. The final grid used consists of about 1.4 million models based on about 340\,000 original BaSTI models.

\subsection{Testing the method}

To test our algorithm we use test star 6 from \citet{bas10} from their Table 1. The input \num\ and \dnu\ are 21$\pm$0.2 and 2.2$\pm$0.05\mh , respectively, where we adopt typical uncertainties from our sample of red giants. The resulting model probabilities and cumulative probability distributions for the model radius are illustrated in Fig.\,\ref{fig:teststar}, from which we infer the most probable radius and its uncertainty to be R = 22.1$\pm$0.9\,R\sun , which is in good agreement with the input radius of 21.44\,R\sun\ \citep{bas10}.

A more realistic test case is the red giant $\epsilon$ Oph, for which an interferometric radius is available. With \num\ = 53.5$\pm$2.0, \dnu\ = 5.2$\pm$0.1\mh\ \citep{kal08a}, and a spectroscopic \teff\ = 4877$\pm$100\,K \citep{rid06} we determined a radius of 10.7$\pm$0.4\,R\sun , which is in good agreement with the interferometric radius of 10.55$\pm$0.15\,R\sun\ \citep{maz09}. The other parameters are also in fairly good agreement with independent measurements. Our mass estimate of 1.89$\pm$0.08\,M\sun\ compares well to stellar masses determined from a detailed modelling: 1.85$\pm$0.05\,M\sun\ \citep{maz09} and 2.02\,M\sun\ \citep{kal08a}. Even our luminosity estimate of 59$\pm$5\,L\sun\ is compatible with the luminosity of 58$\pm$4L\sun\ based on the Hipparcos parallax \citep{vanLee07}.

\subsection{Results for \textit{Kepler} stars}

In a first step, we have excluded 65 red giants which are associated with clusters from our sample because they might bias the subsequent analysis. Fig.\,\ref{fig:RMcomp} shows the radius as a function of the mass for the remaining sample of red giants. Whereas in the top panel, $R$ and $M$ are directly determined from the seismic scaling relations adopting the KIC temperatures, the bottom panel shows $R$ and $M$ as they follow from the Bayesian comparison with the stellar model grid. Both distributions include many stars located in a narrow range around ten solar radii. Most of the stars in this range are expected to correspond to the red clump \citep[e.g.,][]{mig09} and their large number is due to the different evolutionary rates of giant-branch stars. Whereas stellar evolution takes place quite rapidly during the RGB phase and after the He flash at the tip of the RGB, it significantly slows down during the quiescent He-core burning phase in the red clump. Hence one can expect to find more red-clump stars and therefore stars with a similar radius than RGB stars in a random sample of red giants. 

Although the two methods to determine $R$ and $M$ give a similar picture there are some important differences. Firstly, the very-low-mass stars in the top panel are most likely artefacts because, apart from binary stars which have lost a significant fraction of their mass, the universe is not old enough for 0.5-0.7 solar mass stars to have become 10-15 solar radii giants. Secondly, the distribution of stars below the red clump seems to be more realistic in the bottom panel than in the top panel. The higher mass stars evolve faster than the lower mass stars and therefore fewer higher mass stars should be present in a random sample of giants. Additionally, the error bars are in general smaller in the bottom panel and the red-clump feature is more pronounced. We are therefore confident that the additional efforts in determining the fundamental parameters are justified because they enable a more detailed picture of the stellar population on the giant branch.

\begin{figure}[b]
	\begin{center}
	\includegraphics[width=0.5\textwidth]{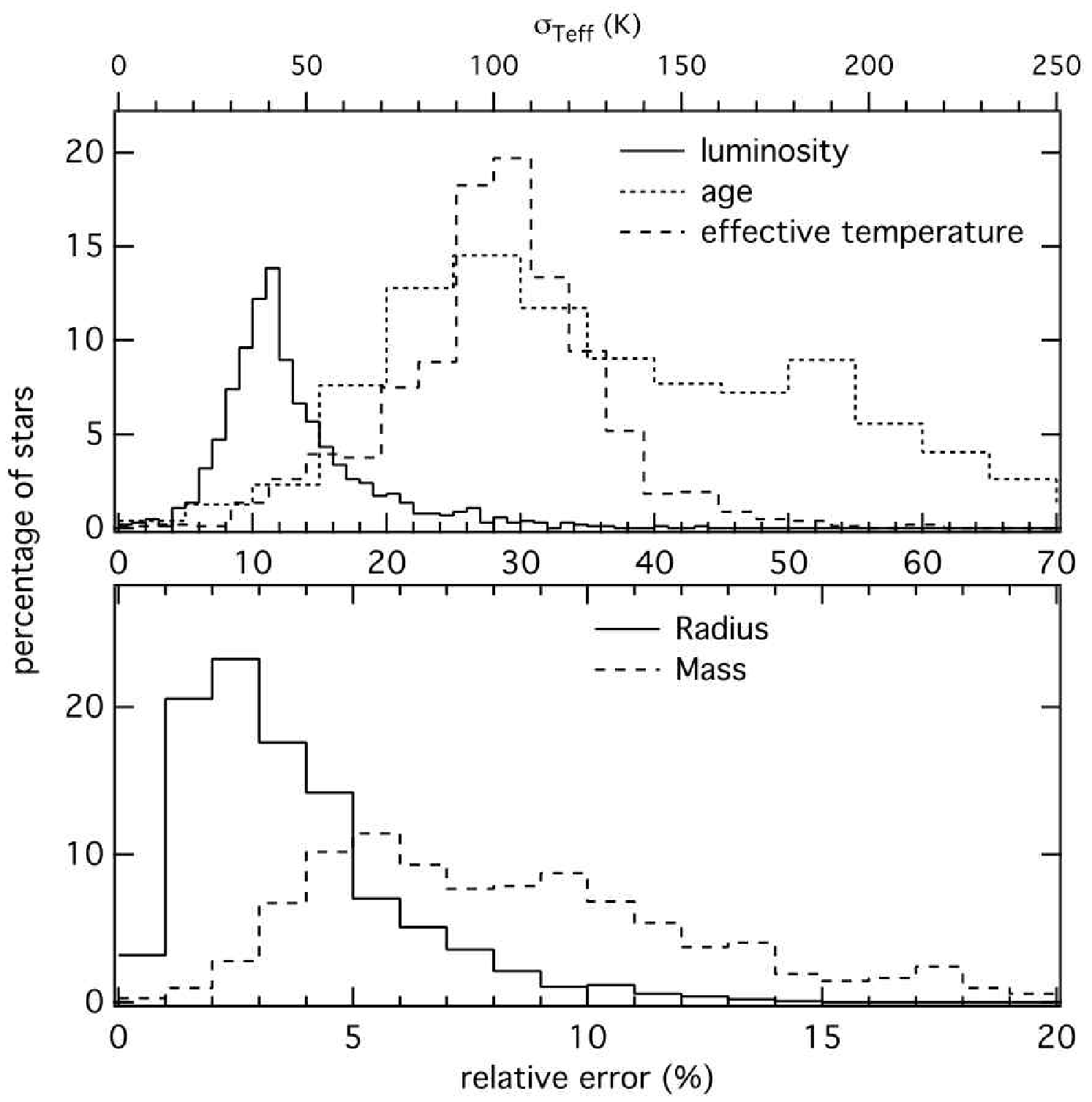}
	\caption{Histograms of the relative (and absolute for \teff ) uncertainties for the asteroseismic fundamental parameter.} 
	\label{fig:RMerror} 
	\end{center} 
\end{figure}

\subsubsection{\textit{Kepler} red giants in the H-R diagram}

As our algorithm also delivers effective temperatures and luminosities we can put the analysed stars in a H-R diagram. Panel $a$ of Fig.\,\ref{fig:HRDcomp} shows the model-independent positions in the H-R diagram, with the temperatures from the KIC and the luminosities following from the Stefan-Boltzmann law, with $R$ directly determined from the seismic scaling relations (by using the KIC temperatures). The other panels show the positions in the H-R diagram as they follow from the Bayesian comparison with the stellar model grid with the colour code indicating the mass, radius, age, and metallicity of the best fit models.   
The distribution of stars in the H-R diagram reveals some interesting features. The most distinctive one (denoted as ``A'' in panel $b$ of Fig.\,\ref{fig:HRDcomp}) consists of a large number of stars ($\sim$35\% of the total sample) lining up at almost constant luminosity and corresponds to the red clump. A similar feature (B) is located at slightly lower luminosities and with an almost constant temperature. It most likely corresponds to the so-called red bump, which is another characteristics of stellar evolution on the giant branch due to the outward-moving hydrogen-burning shell that burns through the mean molecular weight discontinuity left by the first dredge-up from the convective envelope causing a slight contraction of the star before the star starts to ascend the giant branch again. Significantly less populated than the red clump and bump is the feature (C) that corresponds to the so-called secondary clump \citep{gir99} including high-mass stars ($>$2\,M\sun ) that settle as He-core burning stars at lower luminosities than the lower-mass red-clump stars. The secondary-clump population is of particular interest because its mass puts tight constraints on, e.g., the convective-core overshooting or the recent history of star formation in the Galaxy \citep{gir99}. We refer to \citet{hub10} who report on the signature of secondary-clump stars in the distributions of the seismic observables of red giants in the \textit{Kepler} field of view (see also \citealt{mos10} for CoRoT red giants).

Combining the mass distribution in the H-R diagram (panel $b$ in Fig\,\ref{fig:HRDcomp}) with the radius distribution (panel $c$ in Fig\,\ref{fig:HRDcomp}) we infer that the red clump consists of about 0.8 to 1.8\,M\sun\ stars (with the low-mass stars accumulated at the bottom of the red clump) with a radius between about 10 to 12\,R\sun . The red bump, on the other hand, is dominated by 1\,M\sun\ stars but with a lower average radius than the red-clump stars. The secondary-clump covers a similar radius range  but includes stars with masses above about 2\,M\sun .  Outside the red clumps and bump, the stars follow the usual mass distribution with increasing masses towards higher temperatures and luminosities. The detailed structure in the H-R diagram is almost not visible in the model-independent approach (panel $a$). The uncertain KIC temperatures obviously blur the distribution of the stars in the red clumps and bumps. However, we note that the model-independent approach tentatively shows the mass gradient in the red clump as well.

Although we cannot constrain the metallicity to better than about $\pm$50\% for individual stars, the metallicity distribution shows some interesting trends (see panel $e$ in Fig.\,\ref{fig:HRDcomp}). In the red clump, there is a metallicity gradient ranging from metal-poor stars at the bottom to metal-rich stars at higher luminosities. And the red bump tentatively consists of metal-enhanced stars.

\begin{figure}[t]
	\begin{center}
	\includegraphics[width=0.5\textwidth]{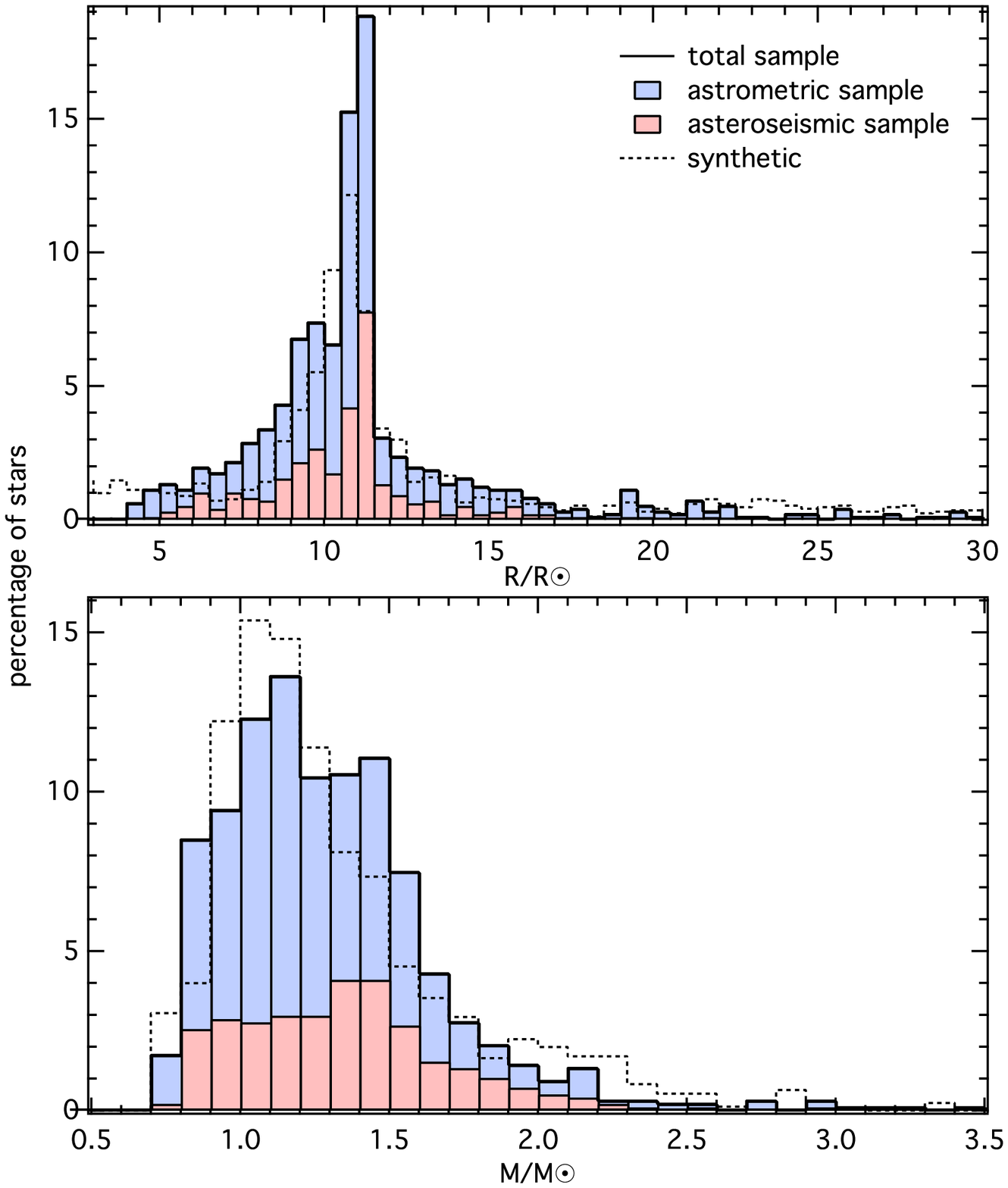}
	\caption{Histograms for the radius and mass distribution in our sample of red-giant stars compared to the distribution of a synthetic red-giant population.} 
	\label{fig:RMhisto} 
	\end{center} 
\end{figure}

To analyse the red clump and bump in more detail we compare in Fig.\,\ref{fig:HRD_Z} the observed features with different metallicity grids. As indicated in panel $d$ of Fig.\,\ref{fig:HRDcomp}, the red clump is dominated by He-core burning stars of the solar-metallicity grid (middle panel of Fig.\,\ref{fig:HRD_Z}). The metallicity gradient in the red clump is due to the position of the He-core burning main-sequence in the different metallicity grids. Whereas the metal-poor models are shifted towards higher temperatures and therefore towards the bottom of the observed red clump, the metal-rich models accumulate at the opposite side of the red clump. Similar can be seen for the red bump. But there, metal-enhanced models are more consistent with the observed sample than solar-metallicity models. 

In Fig.\,\ref{fig:RMerror} we show histograms for the relative uncertainties in mass and radius (bottom panel) showing that for about half of our sample we can constrain the mass and radius to within 7\%  and 3\%, respectively. These uncertainties are dominated by the observational uncertainties of the seismic input parameters and are only slightly affected by the model related parameters. On the other hand, the effective temperature and luminosity of a model with a given mass and radius can significantly differ for, e.g., a different initial chemical composition or mixing length parameter. Compared to Paper I, where only a single metallicity grid was used, we cover a wide range in chemical composition and therefore expect to get a more realistic uncertainty for the H-R diagram position. Histograms for the relative uncertainties in effective temperature, age, and luminosity are given in the top panel of Fig.\,\ref{fig:RMerror} showing that we can determine the effective temperature, age and luminosity to within $\pm$110\,K, $\pm$30\%, and $\pm$11\%, respectively, for most of the analysed red giants.

We have tested our results with the RADIUS \citep{ste09b} and A2Z \citep{mat10} methods, which showed qualitative agreement. However, a direct comparison is meaningless since RADIUS and A2Z use only hydrogen burning models and therefore do not include the red clump.

\subsubsection{Radius and mass distributions}

The radius and mass distribution in our sample of red giants is given in Fig.\,\ref{fig:RMhisto}. The radius distribution clearly shows two populations of stars that are located in the same region of the H-R diagram. The H-shell burning stars ascending the giant branch and the He-core burning stars in the red clump. The very different rate at which they change their fundamental parameters (e.g., the radius) results in two superposed components of their radius distribution. The main component is a broad Gaussian-like distribution with a maximum number of stars between 9.5 and 10\,R\sun . This component is dominated by RGB stars ascending the giant branch and the maximum falls onto the average radius of the red bump (see panel $c$ of Fig.\,\ref{fig:HRDcomp}) but also includes the secondary-clump stars. The RGB stars are superposed with the sharp distribution of red-clump stars with their radii ranging from 10.5 to 11.5\,R\sun .
Also interesting is the excess of stars with a radius around 20\,R\sun . If real, this clustering of stars would be very interesting because it might indicate stars on the AGB whose He-burning shells burn through the discontinuity left by the second dredge up, which happens indeed at about 20 solar radii in solar metallicity models. But a significantly larger sample would be needed to verify if the excess is real. 

The mass distributions in the bottom panel of Fig.\,\ref{fig:RMhisto} shows that there are only very few low-mass stars in our sample. Their number slowly increases between 0.8 to 1.5\,M\sun\ with a small excess between 1 to 1.2\,M\sun , and rapidly drops for higher masses. To test for bias of our composite sample we computed the radius and mass distributions in the subsamples (red and blue bars in Fig.\,\ref{fig:RMhisto}) and found no significant difference for the radius distribution. The mass distributions, on the other hand, are different. Obviously, there are more 1.3 to 1.5\,M\sun\ stars in the asteroseismic sample than in the astrometric sample showing its maximum between 1 to 1.2\,M\sun . We expect the excess of ``high-mass'' stars in the asteroseismic sample to be due to the original selection of the stars. 

The detailed structure in the radius and mass distribution enabled us to identify different stellar populations in our sample which we can compare with synthetic populations for the \textit{Kepler} field of view. To do so, we used the synthetic red-giant population for the \textit{Kepler} field of view  as presented by \citet{hub10} and computed with the stellar synthesis code TRILEGAL \citep{gir05} in the same way as by \citet{mig09} for one of the CoRoT fields. The synthetic radius and mass distributions are indicated as dashed lines in Fig.\,\ref{fig:RMhisto}. Although the comparison can not be done in an absolute sense as the observed sample is biased, the observed and synthetic distributions look quite similar but show also some interesting differences. The red-clump feature in the synthetic radius distribution is slightly broader and has its maximum at a lower radius compared to the observed distribution. Additionally, the RGB component is less pronounced, which is due to significantly less red-bump stars in the synthetic population. Since the stellar synthesis does not include stellar clusters, the main difference in the mass distributions is due to fewer 1.3 to 1.5\,M\sun\ stars in the synthetic distribution. More interesting is, however, the difference for high-mass stars ($>$2\,M\sun ) indicating differently populated secondary clumps. Although these differences potentially carry detailed informations about the star formation history in the \textit{Kepler} field of view and the associated physics, it would require detailed modelling to further investigate them, which is beyond the scope of this paper.

\subsubsection{Testing the KIC parameters}

\begin{figure}[t]
	\begin{center}
	\includegraphics[width=0.5\textwidth]{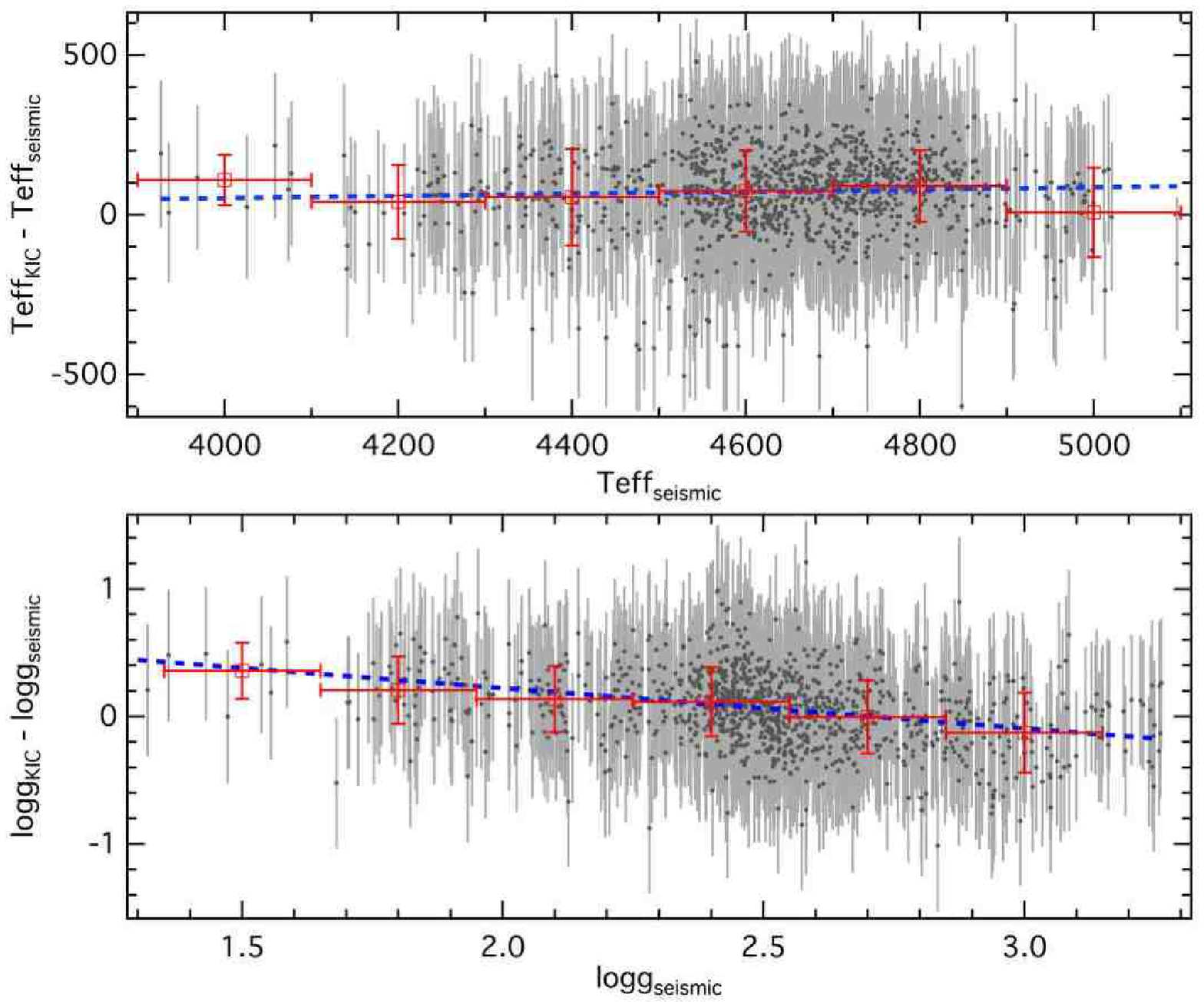}
	\caption{Comparison between the seismically determined effective temperature and surface gravity and the corresponding KIC parameters. Dashed lines indicate a linear fit and red square symbols (in the online version only) show binned values.} 
	\label{fig:Tefflogg} 
	\end{center} 
\end{figure}

Finally, we compared in Fig.\,\ref{fig:Tefflogg} (upper panel) the seismically determined effective temperature with the KIC temperature. We find the seismic temperature systematically shifted towards lower temperatures by about 50\,K (see linear fit and binned values). The rms scatter (about 120\,K) of the temperature difference is, however, within the assumed uncertainties for the KIC temperatures of 200\,K (private communication T. Brown). The KIC uncertainties seem therefore to be overestimated but one has to keep in mind that our algorithm uses the KIC temperatures as an input parameter and the two temperature estimates are therefore not independent. A more meaningful indicator for the reliability of our fundamental parameter estimation is the surface gravity, which is also given in the KIC but not used in our analysis. The seismic mass and radius directly translate to a surface gravity and a comparison between the seismic and KIC surface gravity is shown in the bottom panel of Fig.\,\ref{fig:Tefflogg}. Here we found a systematic difference indicated by a linear fit and binned values. For the red-clump stars (around \lg\ = 2.5) the difference is negligible but drifts apart for stars above (towards lower surface gravity) and below (towards higher surface gravity) the red clump. The difference is, however, less than 0.5 dex for the entire range, and therefore within the uncertainties for the KIC parameters. 

We also compared our seismic \lg\ values for a few stars in common with the spectroscopic study of Bruntt et al.\ (2010, A\&A, in preparation). There is very good agreement with a mean difference (spectroscopy minus seismic \lg ) of $\Delta \log g = +0.03\pm0.15$. We quote the RMS scatter for the seven stars in common.

\section{Summary and conclusions}

We have analysed high-precision photometric time series from the first four months of \textit{Kepler} observations and found more than 1\,000 stars that show a clear power excess in a frequency range typical for solar-type oscillations in red giants. We have applied robust and automated methods to accurately determine the global seismic parameters \num\ and \dnu , and provide an automated identification of the mode degree as well as small spacings for about one half of our sample. We have analysed the uncertainties in our parameter determination and find a clear relation for the uncertainty of \num\ depending on the frequency resolution but also on the height-to-background ratio and the width of the power excess, which are largely determined by the star itself as long as the power excess is well above the white noise of the observations. We have applied our methods to solar data to determine solar reference values for \num\ and \dnu . The measured seismic parameters were then compared to those of an extensive multi-metallicity grid of stellar models, where the seismic parameters of the models were determined from scaling relations. A Bayesian approach for the search of a best fit between observed and model parameters allowed us to derive realistic uncertainties for all fundamental parameters. In principle, we could have estimated the fundamental parameters from the seismic scaling relations using the KIC temperatures. However, we found strong indications that our analysis produced more accurate fundamental parameters and gives us a more detailed view of the stellar populations in our sample of stars.

We have placed the analysed stars in a H-R diagram and found clear features in the distribution of the stars, which we identified as the red clump, the secondary clump, and the red bump. We found a mass gradient in the red clump with the low-mass stars accumulated at the bottom of the red clump and that the red bump is dominated by 1\,M\sun\ stars. Although we cannot determine the chemical composition reliably of individual stars we can conclude in a statistical sense that most of the red-clump stars in our sample are similar to the Sun in terms of the their initial chemical composition with some indications for a metallicity gradient that follows the mass gradient of the red clump. On the other hand we found that the bump stars are more consistent with metal-enhanced stars, which is surprising for a sample of stars that is selected according to criteria that do not constrain the chemical composition. A possible explanation could be that the mixing length parameter used to construct the models is too high. A slightly less efficient convection would shift the solar metallicity red-bump models towards lower temperatures in the direction of the observed red bump. 

The large sample allowed us to investigate the detailed structure of the radius and mass distribution of red giants in the \textit{Kepler} field of view clearly showing the different populations. A comparison with synthetic distributions indicated quantitative agreement but needs further investigations. With the present study and what was presented by \citet{hub10} and \citet{mos10} there are now detailed red-giant populations available for three different regions in the sky, which should be used for future detailed population synthesis studies as first done by \citet{mig09}.

Finally, we mention that although the parameters uncertainties in our analysis are already quite realistic they still represent only a lower limit. There are several effects we do not yet take into account. For example, the frequency dependence of the large frequency spacing \citep[cf][]{mos10} might play a role, as might the weak asymmetry of the power excess humps. To investigate this in more detail we have to wait until \textit{Kepler} can provide significantly longer time series. On the other hand, we expect a larger effect from the limitations of the used model grid. Although our grid covers a wide range in chemical composition, which is one of the parameters that can significantly influence the $M$-$L$-$R$-$T_\mathrm{eff}$ relation, there are other model parameters such as overshooting or a better description of convection, that could change this relation as well and therefore could have consequences to our analysis. These effects are difficult to estimate and are still largely unexplored territory.

For the individual seismic parameters we refer to \citet{hek10b} providing an online table for all red giants observed with \textit{Kepler}. However, our fundamental parameter estimates are not included in this table because they will continuously be improved with the ongoing observations. But we encourage everybody who is interested in our results to request them personally from the lead author.

\begin{acknowledgements}
Funding for the Kepler Mission is provided by NASA's Science Mission Directorate. TK is supported by the Canadian Space Agency, the Austrian Research Promotion Agency, and the Austrian Science Fund. SH, YPE and WJC acknowledge support by the UK Science and Technology Facilities Council. DH acknowledges support by the Astronomical Society of Australia. The research leading to these results has received funding from the European Research Council under the European Community's Seventh Framework Programme (FP7/2007--2013)/ERC grant agreement n$^\circ$227224 (PROSPERITY), as well as from the Research Council of K.U.Leuven grant agreement GOA/2008/04. The National Center for Atmospheric Research is a federally funded research and development center sponsored by the U.S. National Science Foundation. We acknowledge support from the Australian Research Council. The authors gratefully acknowledge the Kepler Science Team and all those who have contributed to making the Kepler Mission possible.
\end{acknowledgements}

\bibliographystyle{aa}
\bibliography{15263}

\begin{thebibliography}{60}
\expandafter\ifx\csname natexlab\endcsname\relax\def\natexlab#1{#1}\fi

\bibitem[{{Ballot} {et~al.}(2006){Ballot}, {Garc{\'{\i}}a}, \&
  {Lambert}}]{bal06}
{Ballot}, J., {Garc{\'{\i}}a}, R.~A., \& {Lambert}, P. 2006, MNRAS, 369, 1281

\bibitem[{{Barban} {et~al.}(2007){Barban}, {Matthews}, {De Ridder}, {Baudin},
  {Kuschnig}, {Mazumdar}, {Samadi}, {Guenther}, {Moffat}, {Rucinski},
  {Sasselov}, {Walker}, \& {Weiss}}]{bar07}
{Barban}, C., {Matthews}, J.~M., {De Ridder}, J., {et~al.} 2007, A\&A, 468,
  1033

\bibitem[{{Basu} {et~al.}(2010){Basu}, {Chaplin}, \& {Elsworth}}]{bas10}
{Basu}, S., {Chaplin}, W.~J., \& {Elsworth}, Y. 2010, ApJ, 710, 1596

\bibitem[{{Batalha} {et~al.}(2010){Batalha}, {Borucki}, {Koch}, {Bryson},
  {Haas}, {Brown}, {Caldwell}, {Hall}, {Gilliland}, {Latham}, {Meibom}, \&
  {Monet}}]{bat10}
{Batalha}, N.~M., {Borucki}, W.~J., {Koch}, D.~G., {et~al.} 2010, ApJL, 713,
  L109

\bibitem[{{Bedding} {et~al.}(2010){Bedding}, {Huber}, {Stello}, {Elsworth},
  {Hekker}, {Kallinger}, {Mathur}, {Mosser}, {Preston}, {Ballot}, {Barban},
  {Broomhall}, {Buzasi}, {Chaplin}, {Garc{\'{\i}}a}, {Gruberbauer}, {Hale}, {De
  Ridder}, {Frandsen}, {Borucki}, {Brown}, {Christensen-Dalsgaard},
  {Gilliland}, {Jenkins}, {Kjeldsen}, {Koch}, {Belkacem}, {Bildsten}, {Bruntt},
  {Campante}, {Deheuvels}, {Derekas}, {Dupret}, {Goupil}, {Hatzes}, {Houdek},
  {Ireland}, {Jiang}, {Karoff}, {Kiss}, {Lebreton}, {Miglio}, {Montalb{\'a}n},
  {Noels}, {Roxburgh}, {Sangaralingam}, {Stevens}, {Suran}, {Tarrant}, \&
  {Weiss}}]{bed10}
{Bedding}, T.~R., {Huber}, D., {Stello}, D., {et~al.} 2010, ApJL, 713, L176

\bibitem[{{Borucki} {et~al.}(2008){Borucki}, {Koch}, {Basri}, {Batalha},
  {Brown}, {Caldwell}, {Christensen-Dalsgaard}, {Cochran}, {Dunham}, {Gautier},
  {Geary}, {Gilliland}, {Jenkins}, {Kondo}, {Latham}, {Lissauer}, \&
  {Monet}}]{bor08}
{Borucki}, W., {Koch}, D., {Basri}, G., {et~al.} 2008, in IAU Symposium, Vol.
  249, IAU Symposium, ed. {Y.-S.~Sun, S.~Ferraz-Mello, \& J.-L.~Zhou}, 17--24

\bibitem[{{Borucki} {et~al.}(2010){Borucki}, {Koch}, {Basri}, {Batalha},
  {Brown}, {Caldwell}, {Caldwell}, {Christensen-Dalsgaard}, {Cochran},
  {DeVore}, {Dunham}, {Dupree}, {Gautier}, {Geary}, {Gilliland}, {Gould},
  {Howell}, {Jenkins}, {Kondo}, {Latham}, {Marcy}, {Meibom}, {Kjeldsen},
  {Lissauer}, {Monet}, {Morrison}, {Sasselov}, {Tarter}, {Boss}, {Brownlee},
  {Owen}, {Buzasi}, {Charbonneau}, {Doyle}, {Fortney}, {Ford}, {Holman},
  {Seager}, {Steffen}, {Welsh}, {Rowe}, {Anderson}, {Buchhave}, {Ciardi},
  {Walkowicz}, {Sherry}, {Horch}, {Isaacson}, {Everett}, {Fischer}, {Torres},
  {Johnson}, {Endl}, {MacQueen}, {Bryson}, {Dotson}, {Haas}, {Kolodziejczak},
  {Van Cleve}, {Chandrasekaran}, {Twicken}, {Quintana}, {Clarke}, {Allen},
  {Li}, {Wu}, {Tenenbaum}, {Verner}, {Bruhweiler}, {Barnes}, \& {Prsa}}]{bor10}
{Borucki}, W.~J., {Koch}, D., {Basri}, G., {et~al.} 2010, Science, 327, 977

\bibitem[{{Broomhall} {et~al.}(2009){Broomhall}, {Chaplin}, {Davies},
  {Elsworth}, {Fletcher}, {Hale}, {Miller}, \& {New}}]{bro09}
{Broomhall}, A., {Chaplin}, W.~J., {Davies}, G.~R., {et~al.} 2009, MNRAS, 396,
  L100

\bibitem[{{Brown} {et~al.}(1991){Brown}, {Gilliland}, {Noyes}, \&
  {Ramsey}}]{bro91}
{Brown}, T.~M., {Gilliland}, R.~L., {Noyes}, R.~W., \& {Ramsey}, L.~W. 1991,
  ApJ, 368, 599

\bibitem[{{Buzasi} {et~al.}(2000){Buzasi}, {Catanzarite}, {Laher}, {Conrow},
  {Shupe}, {Gautier}, {Kreidl}, \& {Everett}}]{buz00}
{Buzasi}, D., {Catanzarite}, J., {Laher}, R., {et~al.} 2000, ApJL, 532, L133

\bibitem[{{Carrier} {et~al.}(2010){Carrier}, {De Ridder}, {Baudin}, {Barban},
  {Hatzes}, {Hekker}, {Kallinger}, {Miglio}, {Montalb{\'a}n}, {Morel}, {Weiss},
  {Auvergne}, {Baglin}, {Catala}, {Michel}, \& {Samadi}}]{car10}
{Carrier}, F., {De Ridder}, J., {Baudin}, F., {et~al.} 2010, A\&A, 509, A73+

\bibitem[{{Christensen-Dalsgaard}(2004)}]{jcd04}
{Christensen-Dalsgaard}, J. 2004, Solar Physics, 220, 137

\bibitem[{{Christensen-Dalsgaard} {et~al.}(2010){Christensen-Dalsgaard},
  {Kjeldsen}, {Brown}, {Gilliland}, {Arentoft}, {Frandsen}, {Quirion},
  {Borucki}, {Koch}, \& {Jenkins}}]{jcd10}
{Christensen-Dalsgaard}, J., {Kjeldsen}, H., {Brown}, T.~M., {et~al.} 2010,
  ApJL, 713, L164

\bibitem[{{Creevey} {et~al.}(2007){Creevey}, {Monteiro}, {Metcalfe}, {Brown},
  {Jim{\'e}nez-Reyes}, \& {Belmonte}}]{cre07}
{Creevey}, O.~L., {Monteiro}, M.~J.~P.~F.~G., {Metcalfe}, T.~S., {et~al.} 2007,
  ApJ, 659, 616

\bibitem[{{De Ridder} {et~al.}(2009){De Ridder}, {Barban}, {Baudin}, {Carrier},
  {Hatzes}, {Hekker}, {Kallinger}, {Weiss}, {Baglin}, {Auvergne}, {Samadi},
  {Barge}, \& {Deleuil}}]{rid09}
{De Ridder}, J., {Barban}, C., {Baudin}, F., {et~al.} 2009, Nature, 459, 398

\bibitem[{{De Ridder} {et~al.}(2006){De Ridder}, {Barban}, {Carrier},
  {Mazumdar}, {Eggenberger}, {Aerts}, {Deruyter}, \& {Vanautgaerden}}]{rid06}
{De Ridder}, J., {Barban}, C., {Carrier}, F., {et~al.} 2006, A\&A, 448, 689

\bibitem[{{Dupret} {et~al.}(2009){Dupret}, {Belkacem}, {Samadi}, {Montalban},
  {Moreira}, {Miglio}, {Godart}, {Ventura}, {Ludwig}, {Grigahc{\`e}ne},
  {Goupil}, {Noels}, \& {Caffau}}]{dup09}
{Dupret}, M., {Belkacem}, K., {Samadi}, R., {et~al.} 2009, A\&A, 506, 57

\bibitem[{{Edmonds} \& {Gilliland}(1996)}]{edm96}
{Edmonds}, P.~D. \& {Gilliland}, R.~L. 1996, ApJL, 464, L157+

\bibitem[{{Frandsen} {et~al.}(2002){Frandsen}, {Carrier}, {Aerts}, {Stello},
  {Maas}, {Burnet}, {Bruntt}, {Teixeira}, {de Medeiros}, {Bouchy}, {Kjeldsen},
  {Pijpers}, \& {Christensen-Dalsgaard}}]{fra02}
{Frandsen}, S., {Carrier}, F., {Aerts}, C., {et~al.} 2002, A\&A, 394, L5

\bibitem[{{Frohlich} {et~al.}(1997){Frohlich}, {Andersen}, {Appourchaux},
  {Berthomieu}, {Crommelynck}, {Domingo}, {Fichot}, {Finsterle}, {Gomez},
  {Gough}, {Jimenez}, {Leifsen}, {Lombaerts}, {Pap}, {Provost}, {Cortes},
  {Romero}, {Roth}, {Sekii}, {Telljohann}, {Toutain}, \& {Wehrli}}]{fro97}
{Frohlich}, C., {Andersen}, B.~N., {Appourchaux}, T., {et~al.} 1997, Solar
  Physics, 170, 1

\bibitem[{{Gilliland} {et~al.}(2010){Gilliland}, {Brown},
  {Christensen-Dalsgaard}, {Kjeldsen}, {Aerts}, {Appourchaux}, {Basu},
  {Bedding}, {Chaplin}, {Cunha}, {De Cat}, {De Ridder}, {Guzik}, {Handler},
  {Kawaler}, {Kiss}, {Kolenberg}, {Kurtz}, {Metcalfe}, {Monteiro}, {Szab{\'o}},
  {Arentoft}, {Balona}, {Debosscher}, {Elsworth}, {Quirion}, {Stello},
  {Su{\'a}rez}, {Borucki}, {Jenkins}, {Koch}, {Kondo}, {Latham}, {Rowe}, \&
  {Steffen}}]{gil10}
{Gilliland}, R.~L., {Brown}, T.~M., {Christensen-Dalsgaard}, J., {et~al.} 2010,
  PASP, 122, 131

\bibitem[{{Girardi}(1999)}]{gir99}
{Girardi}, L. 1999, MNRAS, 308, 818

\bibitem[{{Girardi} {et~al.}(2005){Girardi}, {Groenewegen}, {Hatziminaoglou},
  \& {da Costa}}]{gir05}
{Girardi}, L., {Groenewegen}, M.~A.~T., {Hatziminaoglou}, E., \& {da Costa}, L.
  2005, A\&A, 436, 895

\bibitem[{{Gizon} \& {Solanki}(2003)}]{giz03}
{Gizon}, L. \& {Solanki}, S.~K. 2003, ApJ, 589, 1009

\bibitem[{{Gruberbauer} {et~al.}(2009){Gruberbauer}, {Kallinger}, {Weiss}, \&
  {Guenther}}]{gru09}
{Gruberbauer}, M., {Kallinger}, T., {Weiss}, W.~W., \& {Guenther}, D.~B. 2009,
  A\&A, 506, 1043

\bibitem[{{Harvey}(1985)}]{har85}
{Harvey}, J. 1985, in ESA Special Publication, Vol. 235, Future Missions in
  Solar, Heliospheric \& Space Plasma Physics, ed. {E.~Rolfe \& B.~Battrick},
  199--+

\bibitem[{{Hekker} {et~al.}(2010{\natexlab{a}}){Hekker}, {Broomhall},
  {Chaplin}, {Elsworth}, {Fletcher}, {New}, {Arentoft}, {Quirion}, \&
  {Kjeldsen}}]{hek10c}
{Hekker}, S., {Broomhall}, A., {Chaplin}, W.~J., {et~al.} 2010{\natexlab{a}},
  MNRAS, 402, 2049

\bibitem[{{Hekker} {et~al.}(2010{\natexlab{b}}){Hekker}, {Debosscher}, {Huber},
  {Hidas}, {De Ridder}, {Aerts}, {Stello}, {Bedding}, {Gilliland},
  {Christensen-Dalsgaard}, {Brown}, {Kjeldsen}, {Borucki}, {Koch}, {Jenkins},
  {Van Winckel}, {Beck}, {Blomme}, {Southworth}, {Pigulski}, {Chaplin},
  {Elsworth}, {Stevens}, {Dreizler}, {Kurtz}, {Maceroni}, {Cardini}, {Derekas},
  \& {Suran}}]{hek10}
{Hekker}, S., {Debosscher}, J., {Huber}, D., {et~al.} 2010{\natexlab{b}}, ApJL,
  713, L187

\bibitem[{{Hekker} {et~al.}(2010{\natexlab{c}}){Hekker}, {Elsworth}, {De
  Ridder}, {Buzasi}, {Huber}, \& {Kallinger}}]{hek10b}
{Hekker}, S., {Elsworth}, Y., {De Ridder}, J., {et~al.} 2010{\natexlab{c}},
  A\&A (submitted)

\bibitem[{{Hekker} {et~al.}(2009){Hekker}, {Kallinger}, {Baudin}, {De Ridder},
  {Barban}, {Carrier}, {Hatzes}, {Weiss}, \& {Baglin}}]{hek09}
{Hekker}, S., {Kallinger}, T., {Baudin}, F., {et~al.} 2009, A\&A, 506, 465

\bibitem[{{Huber} {et~al.}(2010){Huber}, {Bedding}, {Stello}, {Buzasi},
  {Hekker}, \& {Kallinger}}]{hub10}
{Huber}, D., {Bedding}, T.~R., {Stello}, D., {et~al.} 2010, ApJ (submitted)

\bibitem[{{Huber} {et~al.}(2009){Huber}, {Stello}, {Bedding}, {Chaplin},
  {Arentoft}, {Quirion}, \& {Kjeldsen}}]{hub09}
{Huber}, D., {Stello}, D., {Bedding}, T.~R., {et~al.} 2009, Communications in
  Asteroseismology, 160, 74

\bibitem[{{Jenkins} {et~al.}(2010){Jenkins}, {Caldwell}, {Chandrasekaran},
  {Twicken}, {Bryson}, {Quintana}, {Clarke}, {Li}, {Allen}, {Tenenbaum}, {Wu},
  {Klaus}, {Van Cleve}, {Dotson}, {Haas}, {Gilliland}, {Koch}, \&
  {Borucki}}]{jen10}
{Jenkins}, J.~M., {Caldwell}, D.~A., {Chandrasekaran}, H., {et~al.} 2010, ApJL,
  713, L120

\bibitem[{{Kallinger} {et~al.}(2008{\natexlab{a}}){Kallinger}, {Guenther},
  {Matthews}, {Weiss}, {Huber}, {Kuschnig}, {Moffat}, {Rucinski}, \&
  {Sasselov}}]{kal08a}
{Kallinger}, T., {Guenther}, D.~B., {Matthews}, J.~M., {et~al.}
  2008{\natexlab{a}}, A\&A, 478, 497

\bibitem[{{Kallinger} {et~al.}(2008{\natexlab{b}}){Kallinger}, {Guenther},
  {Weiss}, {Hareter}, {Matthews}, {Kuschnig}, {Reegen}, {Walker}, {Rucinski},
  {Moffat}, \& {Sasselov}}]{kal08b}
{Kallinger}, T., {Guenther}, D.~B., {Weiss}, W.~W., {et~al.}
  2008{\natexlab{b}}, CoAst, 153, 84

\bibitem[{{Kallinger} {et~al.}(2010){Kallinger}, {Weiss}, {Barban}, {Baudin},
  {Cameron}, {Carrier}, {De Ridder}, {Goupil}, {Gruberbauer}, {Hatzes},
  {Hekker}, {Samadi}, \& {Deleuil}}]{kal10}
{Kallinger}, T., {Weiss}, W.~W., {Barban}, C., {et~al.} 2010, A\&A, 509, A77+

\bibitem[{{Kallinger} {et~al.}(2005){Kallinger}, {Zwintz}, {Pamyatnykh},
  {Guenther}, \& {Weiss}}]{kal05}
{Kallinger}, T., {Zwintz}, K., {Pamyatnykh}, A.~A., {Guenther}, D.~B., \&
  {Weiss}, W.~W. 2005, A\&A, 433, 267

\bibitem[{{Kjeldsen} \& {Bedding}(1995)}]{kje95}
{Kjeldsen}, H. \& {Bedding}, T.~R. 1995, A\&A, 293, 87

\bibitem[{Kjeldsen {et~al.}(2008)Kjeldsen, Bedding, \&
  Christensen-Dalsgaard}]{kje08}
Kjeldsen, H., Bedding, T.~R., \& Christensen-Dalsgaard, J. 2008, ApJ, 683, L175

\bibitem[{{Latham} {et~al.}(2005){Latham}, {Brown}, {Monet}, {Everett},
  {Esquerdo}, \& {Hergenrother}}]{lat05}
{Latham}, D.~W., {Brown}, T.~M., {Monet}, D.~G., {et~al.} 2005, in Bulletin of
  the American Astronomical Society, Vol.~37, Bulletin of the American
  Astronomical Society, 1340--+

\bibitem[{{Mathur} {et~al.}(2010){Mathur}, {Garc{\'{\i}}a}, {R{\'e}gulo},
  {Creevey}, {Ballot}, {Salabert}, {Arentoft}, {Quirion}, {Chaplin}, \&
  {Kjeldsen}}]{mat10}
{Mathur}, S., {Garc{\'{\i}}a}, R.~A., {R{\'e}gulo}, C., {et~al.} 2010, A\&A,
  511, A46+

\bibitem[{{Mazumdar} {et~al.}(2009){Mazumdar}, {M{\'e}rand}, {Demarque},
  {Kervella}, {Barban}, {Baudin}, {Coud{\'e} du Foresto}, {Farrington},
  {Goldfinger}, {Goupil}, {Josselin}, {Kuschnig}, {McAlister}, {Matthews},
  {Ridgway}, {Sturmann}, {Sturmann}, {ten Brummelaar}, \& {Turner}}]{maz09}
{Mazumdar}, A., {M{\'e}rand}, A., {Demarque}, P., {et~al.} 2009, A\&A, 503, 521

\bibitem[{{Merline}(1999)}]{mer99}
{Merline}, W.~J. 1999, in Astronomical Society of the Pacific Conference
  Series, Vol. 185, IAU Colloq. 170: Precise Stellar Radial Velocities, ed.
  {J.~B.~Hearnshaw \& C.~D.~Scarfe}, 187--+

\bibitem[{{Miglio} {et~al.}(2009){Miglio}, {Montalb{\'a}n}, {Baudin},
  {Eggenberger}, {Noels}, {Hekker}, {De Ridder}, {Weiss}, \& {Baglin}}]{mig09}
{Miglio}, A., {Montalb{\'a}n}, J., {Baudin}, F., {et~al.} 2009, A\&A, 503, L21

\bibitem[{{Monet} {et~al.}(2010){Monet}, {Jenkins}, {Dunham}, {Bryson},
  {Gilliland}, {Latham}, {Borucki}, \& {Koch}}]{mon10}
{Monet}, D.~G., {Jenkins}, J.~M., {Dunham}, E.~W., {et~al.} 2010, ArXiv
  e-prints

\bibitem[{{Mosser} \& {Appourchaux}(2009)}]{mos09}
{Mosser}, B. \& {Appourchaux}, T. 2009, A\&A, 508, 877

\bibitem[{{Mosser} {et~al.}(2010){Mosser}, {Belkacem}, {Goupil}, {Miglio},
  {Morel}, {Barban}, {Baudin}, {Hekker}, {Samadi}, {De Ridder}, {Weiss},
  {Auvergne}, \& {Baglin}}]{mos10}
{Mosser}, B., {Belkacem}, K., {Goupil}, M.~J., {et~al.} 2010, ArXiv e-prints

\bibitem[{{Nigam} {et~al.}(1998){Nigam}, {Kosovichev}, {Scherrer}, \&
  {Schou}}]{nig98}
{Nigam}, R., {Kosovichev}, A.~G., {Scherrer}, P.~H., \& {Schou}, J. 1998, ApJL,
  495, L115+

\bibitem[{{Pietrinferni} {et~al.}(2004){Pietrinferni}, {Cassisi}, {Salaris}, \&
  {Castelli}}]{pie04}
{Pietrinferni}, A., {Cassisi}, S., {Salaris}, M., \& {Castelli}, F. 2004, ApJ,
  612, 168

\bibitem[{{Retter} {et~al.}(2003){Retter}, {Bedding}, {Buzasi}, {Kjeldsen}, \&
  {Kiss}}]{ret03}
{Retter}, A., {Bedding}, T.~R., {Buzasi}, D.~L., {Kjeldsen}, H., \& {Kiss},
  L.~L. 2003, ApJL, 591, L151

\bibitem[{{Stello} {et~al.}(2010){Stello}, {Basu}, {Bruntt}, {Mosser},
  {Stevens}, {Brown}, {Christensen-Dalsgaard}, {Gilliland}, {Kjeldsen},
  {Arentoft}, {Ballot}, {Barban}, {Bedding}, {Chaplin}, {Elsworth},
  {Garc{\'{\i}}a}, {Goupil}, {Hekker}, {Huber}, {Mathur}, {Meibom},
  {Sangaralingam}, {Baldner}, {Belkacem}, {Biazzo}, {Brogaard}, {Su{\'a}rez},
  {D'Antona}, {Demarque}, {Esch}, {Gai}, {Grundahl}, {Lebreton}, {Jiang},
  {Jevtic}, {Karoff}, {Miglio}, {Molenda-{\.Z}akowicz}, {Montalb{\'a}n},
  {Noels}, {Roca Cort{\'e}s}, {Roxburgh}, {Serenelli}, {Silva Aguirre},
  {Sterken}, {Stine}, {Szab{\'o}}, {Weiss}, {Borucki}, {Koch}, \&
  {Jenkins}}]{ste10}
{Stello}, D., {Basu}, S., {Bruntt}, H., {et~al.} 2010, ApJL, 713, L182

\bibitem[{{Stello} {et~al.}(2007){Stello}, {Bruntt}, {Kjeldsen}, {Bedding},
  {Arentoft}, {Gilliland}, {Nuspl}, {Kim}, {Kang}, {Koo}, {Lee}, {Sterken},
  {Lee}, {Jensen}, {Jacob}, {Szab{\'o}}, {Frandsen}, {Csubry}, {Dind},
  {Bouzid}, {Dall}, \& {Kiss}}]{ste07}
{Stello}, D., {Bruntt}, H., {Kjeldsen}, H., {et~al.} 2007, MNRAS, 377, 584

\bibitem[{{Stello} {et~al.}(2008){Stello}, {Bruntt}, {Preston}, \&
  {Buzasi}}]{ste08}
{Stello}, D., {Bruntt}, H., {Preston}, H., \& {Buzasi}, D. 2008, ApJL, 674, L53

\bibitem[{{Stello} {et~al.}(2009{\natexlab{a}}){Stello}, {Chaplin}, {Basu},
  {Elsworth}, \& {Bedding}}]{ste09c}
{Stello}, D., {Chaplin}, W.~J., {Basu}, S., {Elsworth}, Y., \& {Bedding}, T.~R.
  2009{\natexlab{a}}, MNRAS, 400, L80

\bibitem[{{Stello} {et~al.}(2009{\natexlab{b}}){Stello}, {Chaplin}, {Bruntt},
  {Creevey}, {Garc{\'{\i}}a-Hern{\'a}ndez}, {Monteiro}, {Moya}, {Quirion},
  {Sousa}, {Su{\'a}rez}, {Appourchaux}, {Arentoft}, {Ballot}, {Bedding},
  {Christensen-Dalsgaard}, {Elsworth}, {Fletcher}, {Garc{\'{\i}}a}, {Houdek},
  {Jim{\'e}nez-Reyes}, {Kjeldsen}, {New}, {R{\'e}gulo}, {Salabert}, \&
  {Toutain}}]{ste09b}
{Stello}, D., {Chaplin}, W.~J., {Bruntt}, H., {et~al.} 2009{\natexlab{b}}, ApJ,
  700, 1589

\bibitem[{{Stello} \& {Gilliland}(2009)}]{ste09a}
{Stello}, D. \& {Gilliland}, R.~L. 2009, ApJ, 700, 949

\bibitem[{{Tarrant} {et~al.}(2007){Tarrant}, {Chaplin}, {Elsworth},
  {Spreckley}, \& {Stevens}}]{tar07}
{Tarrant}, N.~J., {Chaplin}, W.~J., {Elsworth}, Y., {Spreckley}, S.~A., \&
  {Stevens}, I.~R. 2007, MNRAS, 382, L48

\bibitem[{{Tassoul}(1980)}]{tas80}
{Tassoul}, M. 1980, ApJS, 43, 469

\bibitem[{{Toutain} \& {Froehlich}(1992)}]{tou92}
{Toutain}, T. \& {Froehlich}, C. 1992, A\&A, 257, 287

\bibitem[{{van Leeuwen}(2007)}]{vanLee07}
{van Leeuwen}, F. 2007, A\&A, 474, 653

\end{thebibliography}

\end{document}